\documentclass[fleqn,usenatbib]{mnras}

\usepackage{txfonts}

\usepackage[T1]{fontenc}

\DeclareRobustCommand{\VAN}[3]{#2}
\let\VANthebibliography\thebibliography
\def\thebibliography{\DeclareRobustCommand{\VAN}[3]{##3}\VANthebibliography}

\usepackage{graphicx}
\usepackage{url}	
\usepackage{longtable}
\usepackage{soul}
\usepackage{textcomp}
\usepackage{caption}
\usepackage{subcaption}
\usepackage{caption}
\usepackage{array}
\usepackage{hyperref}
\usepackage{natbib}

\title[MIGHTEE $\alpha$-$z$ correlation]{MIGHTEE: exploring the relationship between spectral index, redshift and radio luminosity}

\author[S. Pinjarkar et al.]{Siddhant Pinjarkar$^{1}$\thanks{E-mail: s.pinjarkar@herts.ac.uk},
Martin J. Hardcastle$^{1}$,
Dharam V. Lal$^{2}$,
Daniel J.B. Smith$^{1}$,
José Afonso$^{3,4}$,\newauthor
Davi Barbosa$^{3,4}$,
Catherine L. Hale$^{5}$,
Matt J. Jarvis$^{5,6}$,
Sthabile Kolwa$^{7}$,
Eric Murphy$^{8}$,\newauthor 
Mattia Vaccari$^{9,10,11}$, and
Imogen H. Whittam$^{5,6}$\\
\\
$^{1}$Centre for Astrophysics Research, Department of Physics, Astronomy and Mathematics, University of Hertfordshire, College Lane, Hatfield AL10 9AB, UK\\
$^{2}$National Centre for Radio Astrophysics - Tata Institute of Fundamental Research, Post Box 3, Ganeshkhind P.O., Pune 411007, India\\
$^{3}$Instituto de Astrofísica e Ciências do Espaço, Universidade de Lisboa, OAL, Tapada da Ajuda, PT1349-018 Lisbon, Portugal.\\
$^{4}$Departamento de Física, Faculdade de Ciências, Universidade de Lisboa, Edifício C8, Campo Grande, PT1749-016 Lisbon.\\
$^{5}$ Astrophysics, Department of Physics, University of Oxford, Keble Road, Oxford, OX1 3RH, UK\\
$^{6}$Department of Physics and Astronomy, University of the Western Cape, Robert Sobukwe Road, 7535 Bellville, Cape Town, South Africa.\\
$^{7}$Physics Department, University of Johannesburg, 5 Kingsway Ave, Rossmore, Johannesburg 2092, South Africa\\
$^{8}$National Radio Astronomy Observatory, Charlottesville, VA, USA\\
$^{9}$Inter-University Institute for Data Intensive Astronomy, Department of Astronomy, University of Cape Town, 7701 Rondebosch, Cape Town, South Africa\\
$^{10}$Inter-University Institute for Data Intensive Astronomy, Department of Physics and Astronomy, University of the Western Cape, 7535 Bellville, Cape Town, South Africa\\
$^{11}$INAF - Istituto di Radioastronomia, via Gobetti 101, 40129 Bologna, Italy\\
}

\begin{document}
\label{firstpage}
\pagerange{\pageref{firstpage}--\pageref{lastpage}}
\maketitle

\begin{abstract}
It has been known for many years that there is an apparent trend for the spectral index ($\alpha$) of radio sources to steepen with redshift $z$, which has led to attempts to select high-redshift objects by searching for radio sources with steep spectra. In this study we use data from the MeerKAT, LOFAR, GMRT, and uGMRT telescopes, particularly using the MIGHTEE and superMIGHTEE surveys, to select compact sources over a wide range of redshifts and luminosities. We investigate the relationship between spectral index, luminosity and redshift and compare our results to those of previous studies. Although there is a correlation between $\alpha$ and $z$ in our sample for some combinations of frequency where good data are available, there is a clear offset between the $\alpha$-$z$ relations in our sample and those derived previously from samples of more luminous objects; in other words, the $\alpha$-$z$ relation is different for low and high luminosity sources. The relationships between $\alpha$ and luminosity are also weak in our sample but in general the most luminous sources are steeper-spectrum and this trend is extended by samples from previous studies. In detail, we argue that both a $\alpha$-luminosity relation and an $\alpha$-$z$ relation can be found in the data, but it is the former that drives the apparent $\alpha$-$z$ relation observed in earlier work, which only appears because of the strong redshift-luminosity relation in bright, flux density-limited samples. Steep-spectrum selection should be applied with caution in searching for high-$z$ sources in future deep surveys.
\end{abstract}




\section{Introduction}
\label{Intro}
Radio galaxies and radio-loud quasars are believed to be powered by accretion of matter onto the supermassive black holes located at the centre of the host galaxy. Distant radio sources are among the most massive, luminous and largest objects in the universe (see review by \citealt{Mileyetal2008}). These distant objects are called high-redshift radio galaxies (HzRGs) and are known for their kpc-scale jets and lobes, clumpy optical morphology~\citep{Villaretal2003, Reulandetal2003}, and high stellar masses~\citep{Seymouretal2007, Debreucketal2010}. The radio emission from these objects is due to the synchrotron process, in which flat spectral indices ($\alpha \approx -0.5$) are associated with newly accelerated particles and steep spectra ($\alpha \lesssim -1.0$) imply that there have been strong effects of radiative ageing. Selection on the basis of an ultra steep spectrum (USS) has been used to find almost all HzRG, where USS is defined as $\alpha \lesssim -1.4$ and we adopt the convention in which flux density $S_{\nu} \propto \nu^{\alpha}$; we will refer to this relation as the spectral index equation. Often found in protocluster environments~\citep{Pentericcietal2000, Venemansetal2007}, HzRGs have been a topic of interest since the 1960s, when \cite{Minkowski1960} used the association of bright emission lines and the bright radio source 3C 295, to determine its redshift of $z = 0.5$. Since then many HzRG sources or candidates have been identified~\citep{Rawlingsetal1996, Debreucketal1998, Debreucketal1999, Mileyetal2008}; the most recent HzRG to be discovered on the basis of its steep radio spectrum was found by \cite{Saxenaetal2018} at $z = 5.72$ with an ultra-steep spectral index value, $\alpha = -1.4$ between 150 MHz and 1.4 GHz. 

Early studies like those of \cite{Tielensetal1979} and \cite{Blumenthaletal1979} first showed that USS sources had smaller angular sizes, implying that they were located farther away, and also recognized the association of redshift with spectral index. Later, more studies began exploring the underlying causes of this correlation; possibilities included star formation rates, $K$-correction for steep spectral index, luminosity dependency, galaxy cluster richness, etc. (e.g.; \citealt{Chambersetal1990, Kroliketal1991, Athreyaetal1998}, \citealt{Klameretal2006, Vernstrometal2018}). It was realised early on that one plausible reason for the relationship between spectral index and redshift (hereafter the $\alpha$-$z$ relation) was inverse Compton scattering losses, which differentially affect high-$z$ sources due to the higher energy density in cosmic microwave background photons at high $z$ \citep{Kroliketal1991}. \cite{Morabitoetal2018} used models including redshift dependent inverse Compton losses to simulate HzRGs and explore the spectral index redshift correlation, and compared their simulated sample to that of \cite{Debrucketal2000a} who used objects from the 3CR survey \citep{Spinardetal1985} and the Molonglo Reference Catalog (MRC) survey \citep{Mccarthyetal1996}. \cite{Morabitoetal2018} found that the $\alpha$-$z$ correlation existed in their sample, as also reported by \cite{Debrucketal2000a}, and suggested that the spectral index criterion used to find USS can be relaxed to $\alpha < -0.9 $ or $< -0.8$.  \cite{Keretal2012} also confirmed the presence of the $\alpha$-$z$ correlation in high-frequency and low-frequency selected samples, although they do point out that the $\alpha$-$z$ relation is weak and that the intrinsic scatter on $\alpha$ is dominant, arguing that 50\% of the measured gradient was contributed by $K$-correction.

However, investigations carried out by e.g. \cite{Gopalkrishnaetal1988} and \cite{Onuoraetal1989} revealed that a correlation also exists between radio luminosity and spectral index. \cite{Onuoraetal1989} compared different radio luminosity ranges for given redshift bins and observed that with increasing radio luminosity spectral index values become steeper, irrespective of the redshift bin. Similarly, \cite{Gopalkrishnaetal1988} analyzed two samples at 408 MHz of flux densities above 10 Jy and near 1 Jy. They observed different median redshifts for the two samples but similar luminosities and concluded that a spectral index to redshift correlation might not exist for higher redshifts ($z > 1$). Furthermore, the study hints at a correlation between the luminosity and the spectral index of the sample, especially for intermediate-strength source sample (1 Jy sample). \cite{Blundelletal1999}, among others, argue that the luminosity-spectral index correlation is fundamental with the $\alpha$-$z$ correlation being merely a by-product. Clearly, evidence has for some time pointed to the $\alpha$-$z$ correlation taking a secondary role to other correlations with $\alpha$ such as radio luminosity. Some authors have bypassed the ultra-steep spectrum criterion entirely in selecting high-$z$ radio galaxies \citep[e.g.][]{Jarvisetal2009}.

The aim of the current study is to explore the spectral index correlation with redshift and radio luminosity at different frequency ranges using high-sensitivity surveys like MeerKAT International GHz Tiered Extragalactic Exploration (MIGHTEE, \citealt{Jarvisetal2016}) and superMIGHTEE (Lal, Taylor, et al., submitted), which allow us to probe a much wider range of luminosities than earlier studies. With the inception of new and improved telescopes, we can make use of high resolution and high sensitivity surveys to study the relation in greater detail. For the analysis we make use of the first MIGHTEE survey data release 1 \citep{Haleetal2024}. Along with these data, we use the survey carried out by \cite{HaleandJarvisetal2018} using the Low Frequency Array survey (LOFAR), the Giant Meter-wave Radio Telescope survey (GMRT, \citealt{SmolcicSlausandNovaketal2018}) and the early science superMIGHTEE GMRT survey (Lal, Taylor, et al., submitted). 

Using these surveys, we can explore the relation between the spectral index, luminosity and the redshift. Within this paper we therefore aim to answer the following questions using the spectral index analysis at increasing redshift bins for our sample:
\begin{enumerate}[i]
	\item What is the observed relationship between the spectral index and redshift?
	\item What is the observed relationship between the radio luminosity and spectral index?
	\item What is the observed trend for two-frequency and multi-frequency analysis?
	\item Does sample size affect our observations from the two analyses?
	\item What can we infer from the comparison of previous studies to our study?
	\item What are the physical processes that create the spectral index redshift correlation and the radio luminosity spectral index correlation?    
\end{enumerate}

Section 2 describes the data processing steps to get the sample of sources used for our study. In Section 3 we discuss the results obtained from our analysis. The conclusions derived from the analysis are given in Section 4. In this study we use a cosmology in which $H_0 = 70$ km s$^{-1}$ Mpc$^{-1}$, $\Omega_{m}$ = 0.3 and $\Omega_{\Lambda}$ = 0.7.
       
\section{Data reduction and analysis}
\subsection{Data description}
\label{DD}
MIGHTEE \citep{Jarvisetal2016} is providing radio continuum, spectral line, and polarization information for four well-studied extragalactic deep fields: the Cosmological Evolution Survey (COSMOS), \textit{XMM-Newton} Large-Scale Structure (XMM-LSS), Extended Chandra Deep Field-South Survey (ECDFS), and European Large-Area Infrared Space Observatory Survey (ELAIS) S1 fields, using observations with the South African MeerKAT telescope (see \citealt{Jonasetal2009, Jonasetal2016}). MeerKAT is equipped to observe in three bands, namely UHF (544 – 1088 MHz), L-band (856 – 1712 MHz), and S-band (1750 – 3500 MHz). The dense core region of dishes (three-quarters of the collecting area) spans over 1 km in diameter while the rest spreads out to provide a maximum baseline of 8km. The MIGHTEE L-band survey will cover $\sim 20$ deg$^{2}$ over the four extragalactic deep fields at a central frequency of  $\sim 1284$ MHz with $\sim 1000$ h of observations with the L-band receivers. The MIGHTEE data release 1 detects around 70,000 radio sources present in the XMM-LSS field in the form of a catalogue made from images at a resolution of 5 arcsec (see \citealt{Haleetal2024} and \citealt{Heywood2022} for more information about the MIGHTEE survey and details of the data processing steps). The MIGHTEE images at this resolution have a central median rms sensitivity of 3.6 $\mu$Jy beam$^{-1}$.
\begin{figure*}
	\includegraphics[width=4in,height=3.2in]{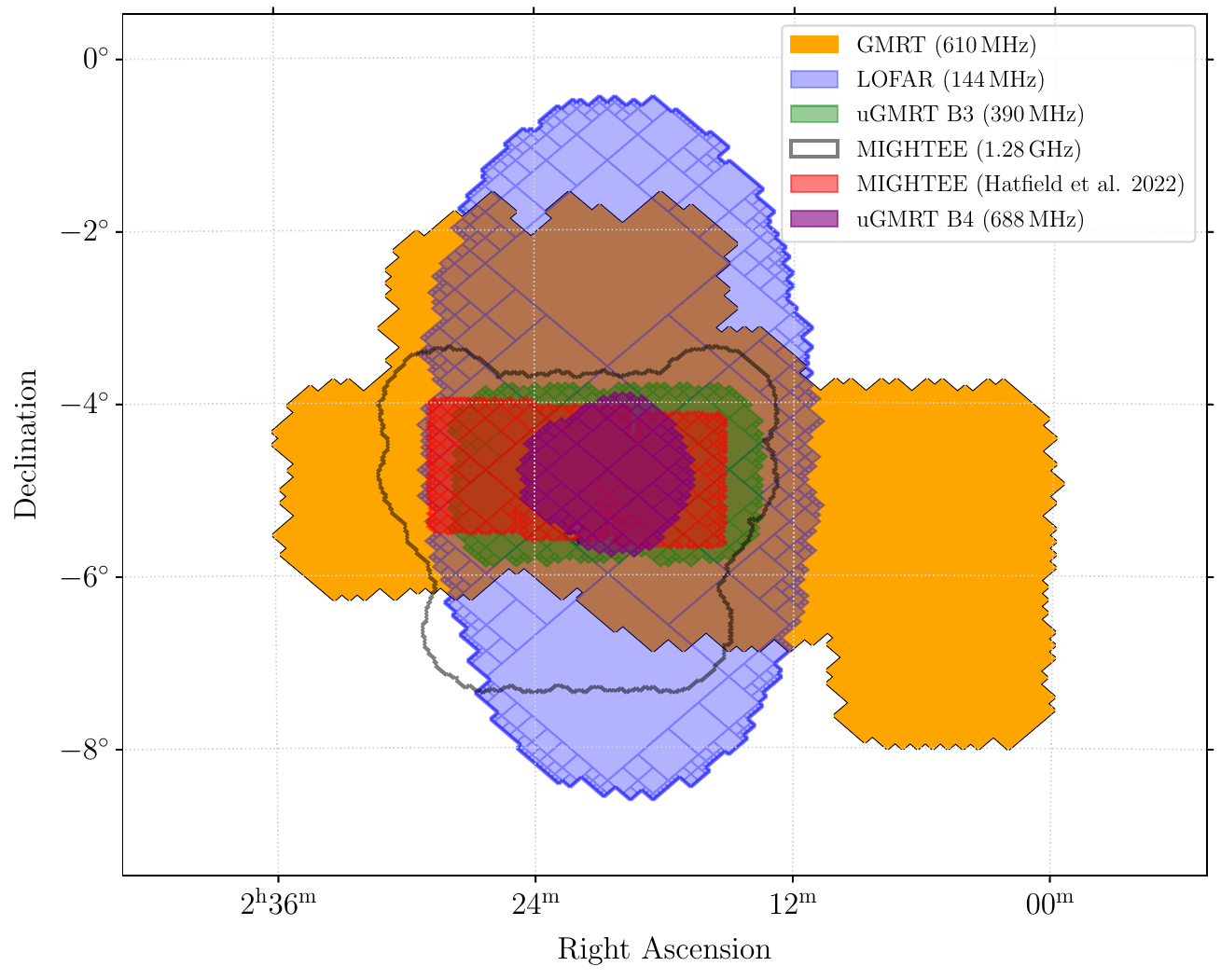}
	\caption{Plot showing the positions of the data from different surveys (see text for details) which will be used in the sample. Note that most of the positions for MIGHTEE are for sources that are matched to the photometric redshift catalogue, and so are limited to the sky coverage of the optical survey \citep{Hatfieldetal2022}. All MIGHTEE positions including those with a spectroscopic catalogue match are within the grey line denoting the MIGHTEE survey coverage.}
	\label{WM}
\end{figure*}

We combine the MIGHTEE data with other observations of the field. LOFAR has made a survey of the XMM-LSS field using the High-Band Array (HBA) at 120-168 MHz which covers almost the entire area currently observed by MIGHTEE, but at significantly lower sensitivity for a typical radio spectral index of $\sim -0.7$. The observations in the field reach a central rms of 280 {$\mu$}Jy beam$^{-1}$ at 144 MHz and provide a resolution of $7.5\times 8.5$ arcsec (\citealt{HaleandJarvisetal2018}). The XMM-LSS was also observed with the GMRT, \cite{SmolcicSlausandNovaketal2018}, in a 610-MHz radio continuum survey covering a 25 deg$^{2}$ area in the XMM-LSS field and towards the XXL-N field. The rms noise level achieved in the XMM-LSS field is between 40 {$\mu$}Jy beam$^{-1}$ and 200 {$\mu$}Jy beam$^{-1}$ and the resolution of the mosaic is around 6.5 arcsec. In addition, we use DR1 maps from the superMIGHTEE GMRT survey (Lal, Taylor, et al., submitted), where the observations target the MIGHTEE XMM-LSS early science region~\citep{Heywood2022}. The region used in this study is covered by a mosaic of 4 pointings at GMRT band 3 and 5 pointings at band 4 with a total solid angle of 6.22 deg$^{2}$ and 2.16 deg$^{2}$, with rms sensitivities of 16 $\mu$Jy beam$^{-1}$ and 8 $\mu$Jy beam$^{-1}$ respectively. The band-3 radio frequency covers the range 300 to 500 MHz and band-4 covers 550-900 MHz, out of which we use broad-band data centred at 390 MHz for band-3 and at 688 MHz for band-4. The resolution of both is 10 arcsec (Lal, Taylor, et al., submitted). Fig. \ref{WM} shows a plot of source positions for the five surveys that will be used in this study, indicating their different sky coverages.

\subsection{Data extraction}
\label{DE}
Due to MIGHTEE's high sensitivity, the data obtained from the survey makes it our obvious choice for selecting the radio sources. An advantage of this is that we have the capacity to detect steep-spectrum sources since the MIGHTEE data are much more sensitive than those at the other bands used here. We restrict our analysis to MIGHTEE sources with catalogued deconvolved major axes less than 10 arcsec, which makes it likely that we will be able to obtain a good optical counterpart by simple positional cross-matching. We selected radio sources in the XMM-LSS field that had redshift information. We used the photometric redshift information in the XMM-LSS, reported by \cite{Hatfieldetal2022} and spectroscopic redshift information reported by \cite{Mattiaetal2015, Mattiaetal2022}, to find redshifts of the sources by cross-matching the photometric and spectroscopic catalogues with the MIGHTEE survey by setting the match radius within 2 arcsec, as the positional accuracy for MIGHTEE sources is expected to be better than this; we verified that the distribution of observed offsets deviates from a Rayleigh distribution for larger offsets than 2 arcsec. The cross-match of MIGHTEE with the two redshift catalogues produced a unified MIGHTEE parent sample. We obtain around 1,870 sources that have spectroscopic redshift values available and for the rest we use the photometric redshifts. The cross-match after the union gives 35,478 sources after filtering sources that do not have any redshift information available. For larger sources, visual inspection would be needed to obtain optical counterparts \citep{Pinjarkaretal2023}. This is approximately 50\% less than the number of sources in the DR1 MIGHTEE survey, as we are limited by the area of coverage of the photometric redshift information given by \cite{Hatfieldetal2022} and, in addition, some radio sources do not have an optical counterpart in either redshift catalogue. There are 34,961 sources from the MIGHTEE DR1 survey within the \cite{Hatfieldetal2022} area coverage, of which 96.8\% have an optical counterpart with a redshift estimate: we are therefore not significantly biased by the missing objects (as also seen for the COSMOS field by \citealt{Whittametal2024}). Our selection of compact radio sources allows us to capture the total flux density of a source in the absence of a catalogue that associates components of extended sources. Most high redshift sources are expected to be compact \citep{Blundelletal1999}, so this limitation should not cause a strong bias in our analysis. We return to the question of the effect of extended sources below, in Section \ref{2FA}. We note that for the faintest sources we may be affected by source blending, in which a catalogued radio source is composed of emission from two unrelated physical objects.

\begin{table*}
    \centering
    \begin{tabular}{c c c c c c}
        \hline
        Survey & Frequency (MHz) & Area (deg$^{2}$) & No. of Sources & RMS Depth ($\mu$Jy/beam)& No. crossmatch\\
        \hline \hline
        MIGHTEE & 1280 & 14.4 & 69,059 & 3.6 & 35,478 \\
        LOFAR & 140 & 27 & 3,200 & 280 & 602 \\
        GMRT & 610 & 25 & 6,570 & 200 & 764 \\
        uGMRT-B3 & 390 & 6.22 & 6,226 & 16 & 3,219 \\
        uGMRT-B4 & 688 & 2.16 & 7,243 & 8 & 4,851 \\
        \hline
    \end{tabular}
    \caption{Survey information used to obtain the sample for the spectral index, redshift, and luminosity analysis. The area column gives the area covered by the survey, next is the number of sources in the catalogue for each survey, where the MIGHTEE crossmatch number is the number of sources obtained after cross-matching with the redshift information as described in the text and for the other surveys the number of crossmatches is the total number of sources obtained for each survey after matching them with the optically crossmatched MIGHTEE data. For MIGHTEE we quote the central frequency, but use the effective frequency for each source in our analysis.}
    \label{Table:SurveyInfo}
\end{table*}

We cross-matched the radio co-ordinates of our sample with optical counterparts and redshifts to the radio co-ordinates of sources in the LOFAR survey, the GMRT survey and the superMIGHTEE uGMRT band-3 and band-4 survey using a cross-match radius of 2 arcsec to ensure consistency with the radio-optical cross-match carried out previously. The number of sources with crossmatches at each frequency is given in Table \ref{Table:SurveyInfo}. The number of sources reported by the LOFAR and GMRT surveys is lower than the other surveys as they are less sensitive. Only a source with at least one detection at one of the other surveys in addition to MIGHTEE will be considered further in this work. We note that as we are selecting only compact sources in MIGHTEE, and the resolution of all the surveys used are within a factor 2 of each other, we should be seeing the same emission at each frequency.

\subsection{Parameters and analysis}
\label{sec:PA}
\begin{figure*}
	\includegraphics[width=3.84in]{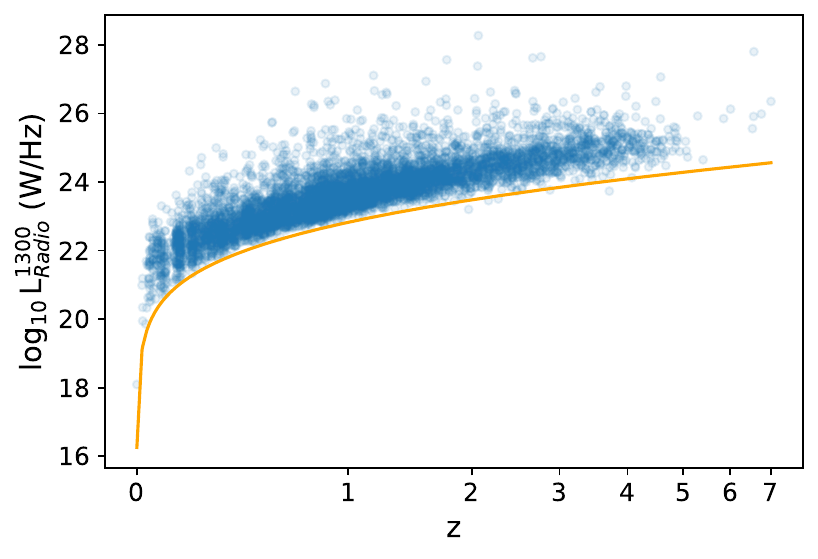}
	\caption{Luminosity versus redshift plot for optically identified MIGHTEE sources with a counterpart at at least one other frequency. The orange line shows the lower limit on luminosity as a function of redshift given the sensitivity of the survey (assuming a $5\sigma$ limit of 18 $\mu$Jy). Note that the objects with $z \ga 6$ are likely photometric redshift outliers \citep{Hatfieldetal2022} and we do not consider them in further analysis. }
	\label{Lz}
\end{figure*}
To find the relationship between the spectral index and redshift for our sample we use the spectral index equation for two given flux densities at two different frequencies. The other variable that we investigate for our sample is the radio luminosity of the sources, computed using the formula
\begin{equation} \label{LR}
	\\L_{\nu_{ref}} = 4\pi{S_{\nu_{obs}}}\left(\frac{\nu_{ref}}{\nu_{obs}}\right)^{\alpha}{D_{L}^{2}}({{1+z})^{-1-\alpha}}
\end{equation}
where, $\nu_{ref}$ is a reference frequency for emission which we take to be 1300 MHz and $\nu_{obs}$ is the effective frequency which varies across the image due to the shape of the primary beam and which we take from the MIGHTEE effective frequency map \citep{Haleetal2024}. $L_{\nu_{ref}}$ is the radio luminosity at 1300 MHz, $S_{\nu_{obs}}$ is the total flux density at a given effective frequency and $D_{L}$ is the luminosity distance. We use the flux densities from MIGHTEE and from the other surveys to evaluate $\alpha$, using the broadest frequency range available for each source, and so calculate the radio luminosities for our sample; this means that our luminosity calculations are not completely homogeneous but they do make use of the best information that we have available for each source. Fig. \ref{Lz} shows the 1300 MHz luminosity of the sample sources as a function of redshift. We can see that the luminosity increases with increasing redshift, as expected, but the scatter in luminosity is large at all redshifts. Thus in this sample the low flux limit allows us to investigate luminosity and redshift dependencies of observed quantities. 

\begin{figure*}
	\includegraphics[width=3.2in,height=2.56in]{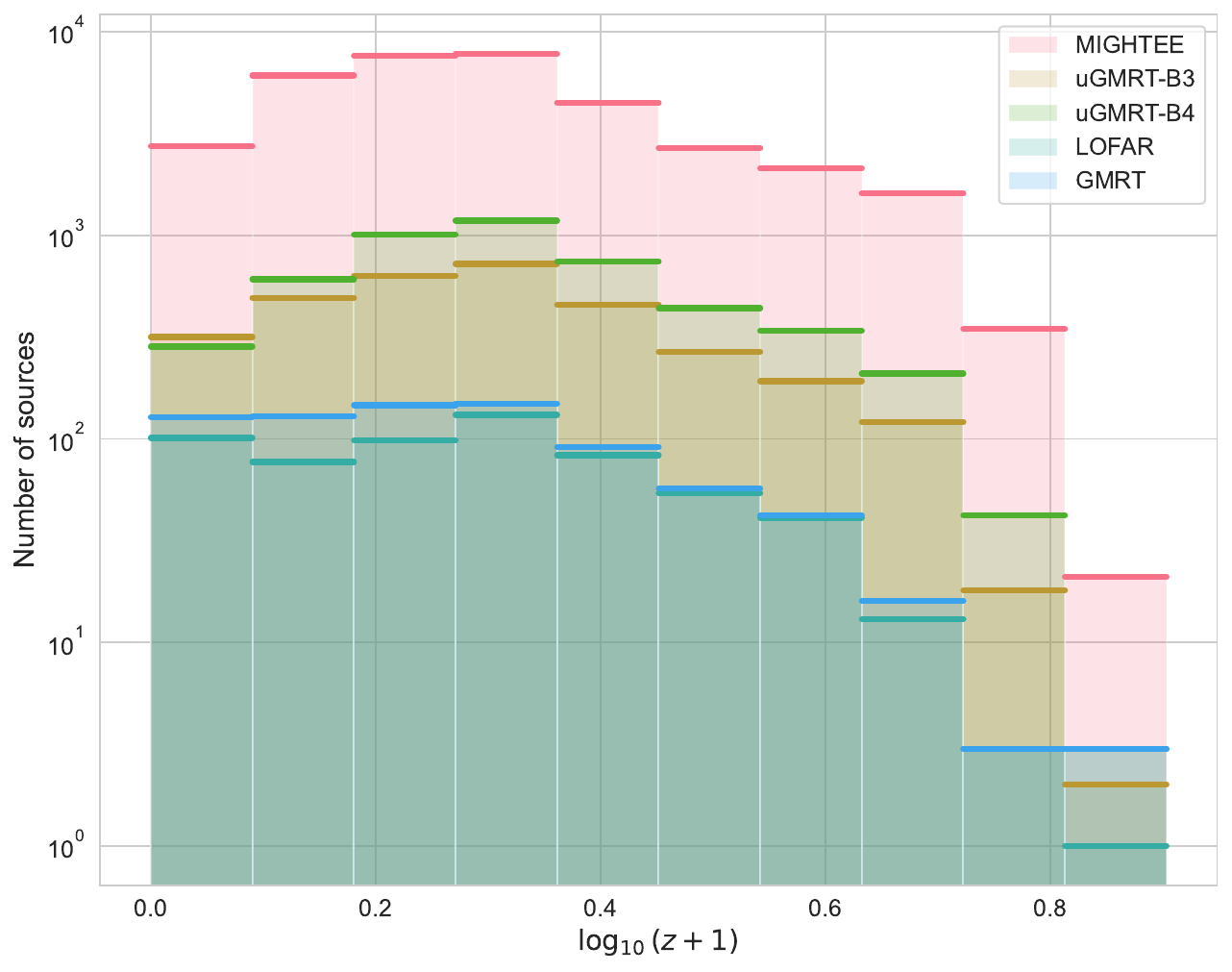}\includegraphics[width=3.2in ,height=2.56in]{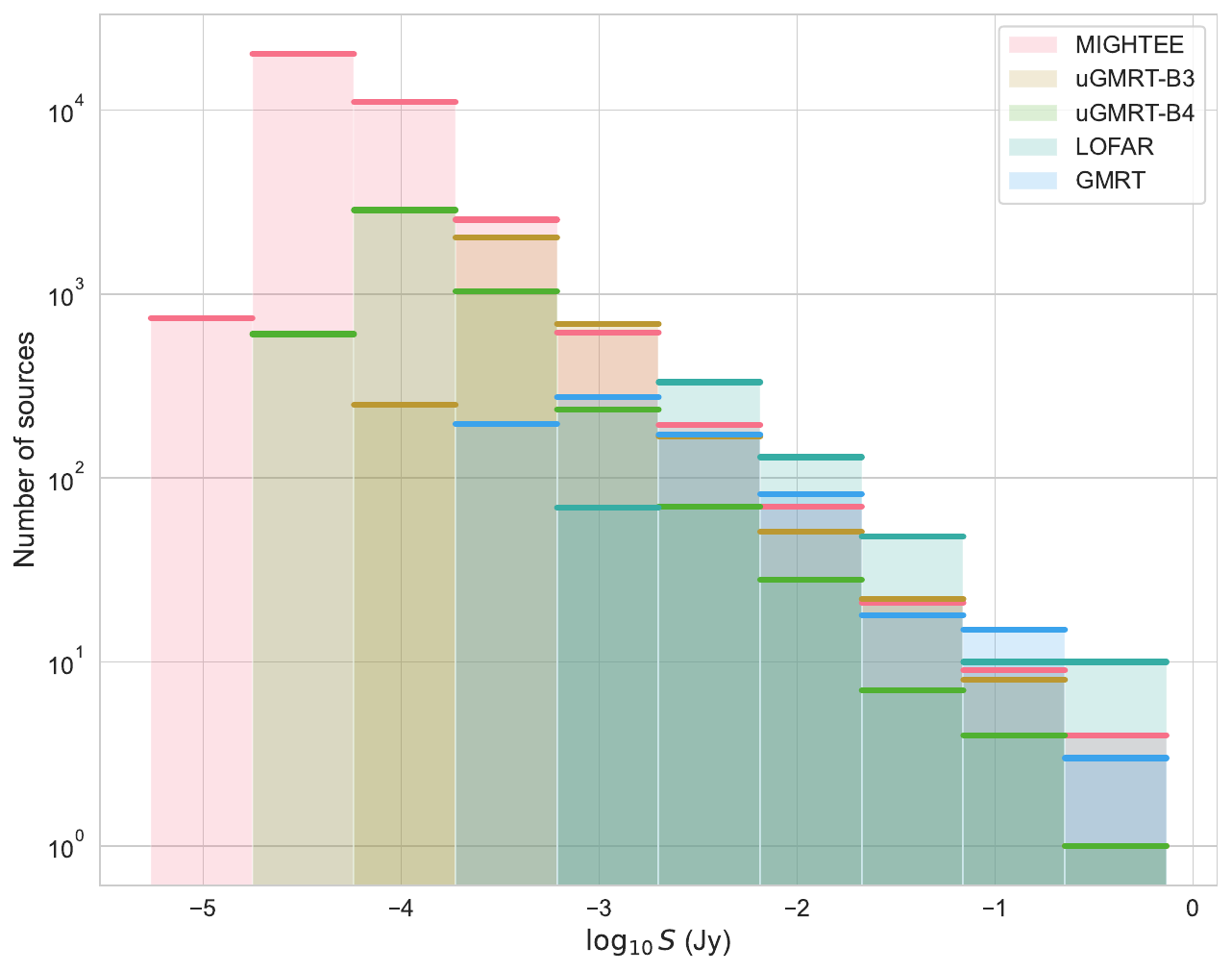}
	\caption{Redshift and flux density distribution (left and right plots respectively) for MIGHTEE survey, LOFAR survey, GMRT survey, uGMRT-B3 survey, and uGMRT-B4 survey.}
	\label{RD}
\end{figure*}

For our parent MIGHTEE sample the redshift values lie between 0 and 7 (see Fig. \ref{RD}, left panel) and more than half of the sources lie within the redshift range of 0-1. The distribution shown in Fig. \ref{RD} (left panel) is obtained for each frequency where the MIGHTEE sources have the highest number of sources as compared to the number of sources with matches in other surveys. In addition, Fig. \ref{RD} (right panel), shows the flux density distribution for the different frequencies, where we can see that the distribution progressively shifts to higher fluxes as we move to lower frequencies. This is the expected behaviour for the dataset, as the sensitivities of the surveys increase with increasing frequency and hence is a good sanity check before we begin the analysis. As various combinations of the LOFAR, the GMRT and the MIGHTEE surveys of the field discussed above will be used further we will refer to the sample with crossmatches in at least one other frequency range as GLaMS (GMRT, LOFAR and MIGHTEE samples).          

It is important to note that our GLaMS sample will contain star-forming galaxies as well as AGN, as we do not have any information that allows us to separate the two classes. At these frequencies both classes of object are dominated by synchrotron radiation and will show qualitatively similar radio spectral indices in many cases. However, above a 1.3 GHz luminosity of $\sim 10^{23}$ W Hz$^{-1}$ we expect AGN to dominate the population \citep{Mauchetal2007, Whittametal2022, Whittametal2024} and above $10^{24}$ W Hz$^{-1}$ there will be essentially no star-forming galaxies \citep{Tadhunteretal2016}. Rather than applying a low-luminosity cutoff, in plots involving luminosity we indicate the position of $L_{1300} = 10^{23}$ W Hz$^{-1}$ so that the reader can be aware of the point below which star-forming galaxies are likely to dominate.

\begin{figure*}
	\includegraphics[width=3.2in]{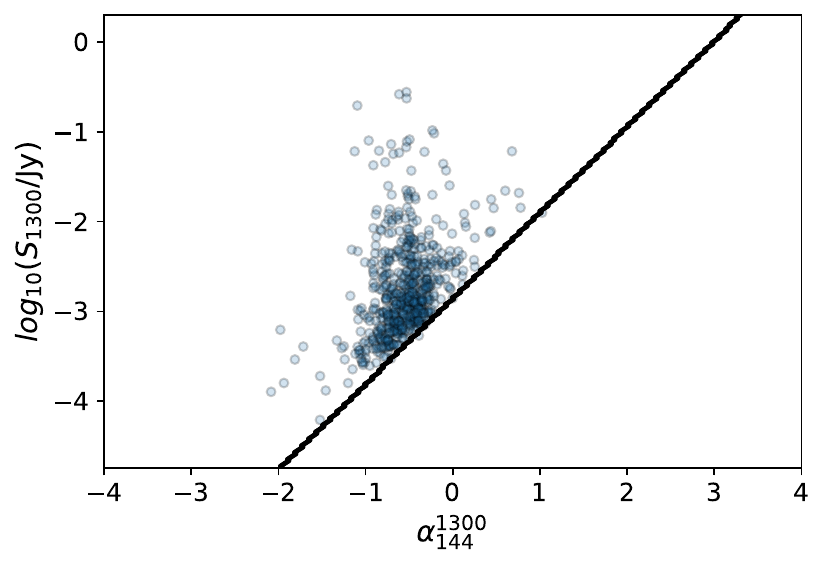}\includegraphics[width=3.2in]{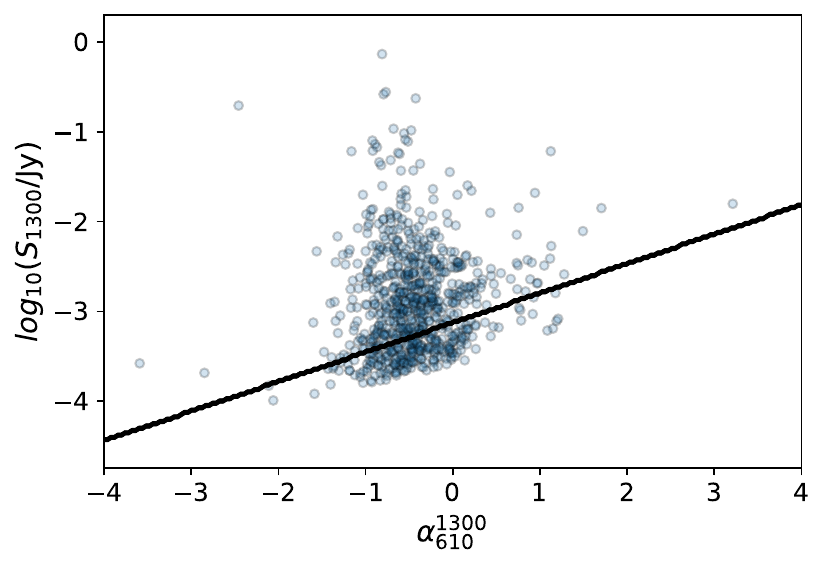}
    \includegraphics[width=3.2in]{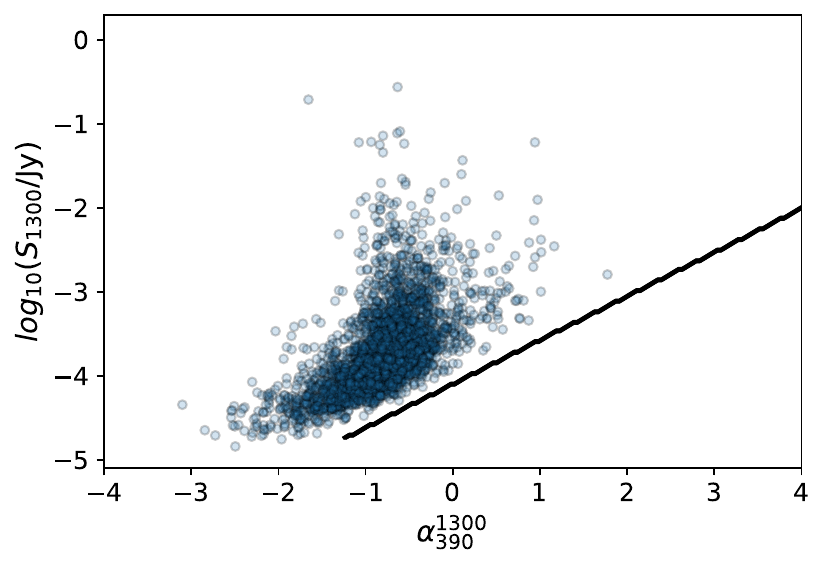}\includegraphics[width=3.2in]{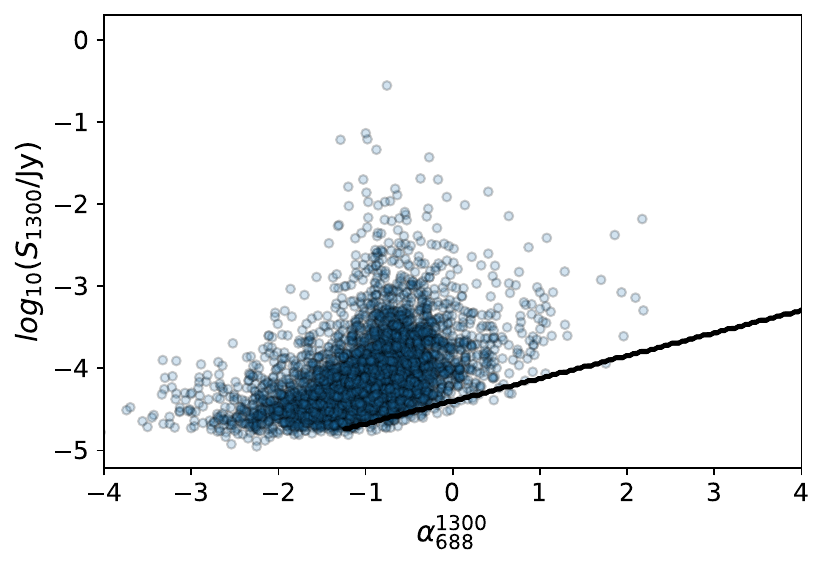}
	\caption{Plots showing the observed distribution of MIGHTEE flux density and spectral index, and the line separating observable from unobservable parts of this parameter space, for  each of the other surveys used. The top left panel shows LOFAR, the top right the GMRT survey, and the bottom panels show the two SuperMIGHTEE bands, 3 and 4 to left and right respectively. The black line is the part of the plot below which sources are excluded by the sensitivity of the corresponding survey. The sources from the GLaMS sample are plotted for their respective frequencies. It can be seen that a broad range of spectral indices can be measured, other than for the faintest sources, and that our principal bias is against inverted-spectrum sources of which there are not many in the parent population (as can be seen from the higher-flux regions of these plots in which the bias is not present).}
	\label{FB}
\end{figure*}

In Fig. \ref{FB}, we show flux density versus spectral index plots where we create a grid that consists of all possible flux densities and spectral indices that can be observed by the MIGHTEE survey. For these points in the grid we evaluate the area under the plot where other frequencies can observe sources observed in the MIGHTEE survey. The solid line in the plot shows the boundary of the undetectable region for the respective survey for the given rms noise of the survey. The plots are overlaid with the sources that are observed in the GLaMS sample where we can see, as expected, that all of the sources lie in the detectable region with the exception of some sources from the GMRT survey. We note that for the GMRT survey the rms noise over the survey area varies from 40 {$\mu$}Jy beam$^{-1}$ to up to 200 {$\mu$}Jy beam$^{-1}$ and as we have used 5$\sigma$ as the detection limit with $\sigma$ being measured in the centre of the GMRT fields for simplicity, some sources are expected to lie in the non-detectable region of the plot due to better sensitivity than we assume. We can see from the plots that we expect to be biased against inverted spectrum sources but we are not biased against steep spectrum sources; moreover, we can see that few observed sources have spectral indices close to the limits imposed by the sensitivity of the survey. This means that we can observe sources that are steep spectrum and faint for the surveys used in this study without worrying about the selection bias; we will argue later that the bias against inverted-spectrum sources does not affect our conclusions. Our only remaining bias is that we only consider relatively compact MIGHTEE sources: thus some extended sources, which might be preferentially steep-spectrum, are excluded from our analysis~\citep{Laingetal1980}. Consideration of these sources requires full optical identification for extended MIGHTEE sources, which is in progress but not available at the time of writing of this paper. However, there is a relatively small number of these sources in the survey \citep{Pinjarkaretal2023}.

Our approach in the remainder of the paper is to consider all of the data without trying to impose any further selection. This has the advantage that we can extract the maximum information from sensitive surveys like the B4 and B3 superMIGHTEE data. Trends seen across many different combinations of frequencies can be considered robust even when the samples considered are not identical.

\subsection{Data from legacy radio surveys}
\label{PSD}
Legacy radio surveys with much higher flux limits than provided by MIGHTEE were the source of the original discovery of the $\alpha$-$z$ relation and so it is important to compare our results with a consistent analysis of objects from those surveys. We use data from the revised Third Cambridge Catalogue of Radio Sources (3CRR;~\citealt{Laingetal1983}) and a subsample of bright sources from the Molonglo Reference Catalog identified with galaxies and quasars \citep{Kapahietal1998a, Kapahietal1998b} to compare with the results obtained from our sample.

The \textsc{3CRR}\footnote{\url{https://3crr.extragalactic.info/field_info.html}} catalog consists of radio sources along with their redshifts, observed by the 3C and Fourth Cambridge Survey (4C) at 178 MHz, along with spectral index, calculated between 178 MHz and 750 MHz. The catalogue contains 173 sources. We use NASA Extragalactic Database (NED) to search sources obtained from the \textsc{3CRR} catalogue. We extract flux densities at 365 MHz by using the Texas survey~\citep{Douglasetal1996}, as the observed frequency for the survey falls in the frequency range we use to get our sample. There are 96 sources in the 3CRR sample with flux densities for frequencies 178 MHz, 365 MHz, and 750 MHz. Similarly, we use the above sample to select sources with flux densities at 1.4 GHz by using the sample obtained by \cite{Paulinyetal1996}, as this frequency is close to the MIGHTEE survey frequency used in our sample\footnote{We use these observations in preference to those from more modern interferometric surveys as they are readily accessible without additional data analysis and are not affected by issues such as missing short spacings or limited surface brightness sensitivity.}. This leaves us with 90 sources in the sample where the source photometric information is available for frequencies at 178 MHz, 365 MHz, 750 MHz, and 1400 MHz. These frequencies are similar to what we use to get our MIGHTEE sample and can be used further to evaluate spectral index and radio luminosities that can be compared with our sample calculations. 

\cite{Kapahietal1998a, Kapahietal1998b} defined a bright subsample of the Molonglo Reference Catalog (MRC) consisting of 557 sources of which 446 are radio galaxies and 111 are radio quasars. A full multi-frequency data compilation for these sources is not available. We therefore obtained photometric data for these sources from NED using the `data products' section given in NASA's ADS website. These data consist of the source names, their redshifts and their respective flux densities at different frequencies. We use these frequencies to obtain the spectral index values for different frequency ranges. This analysis of the 3CRR and MRC sources means that we have the information required to compare the results from our sample and those used by previous studies in a consistent way with frequencies that are matched to those available in GLaMS. The number of MRC sources available depends on the combination of frequencies used but is at most 355.
          
\section{Results and Discussion}
\subsection{Two-frequency analysis}
\label{2FA}
\begin{figure*}
	\includegraphics[width=7.0in,height=7.0in]{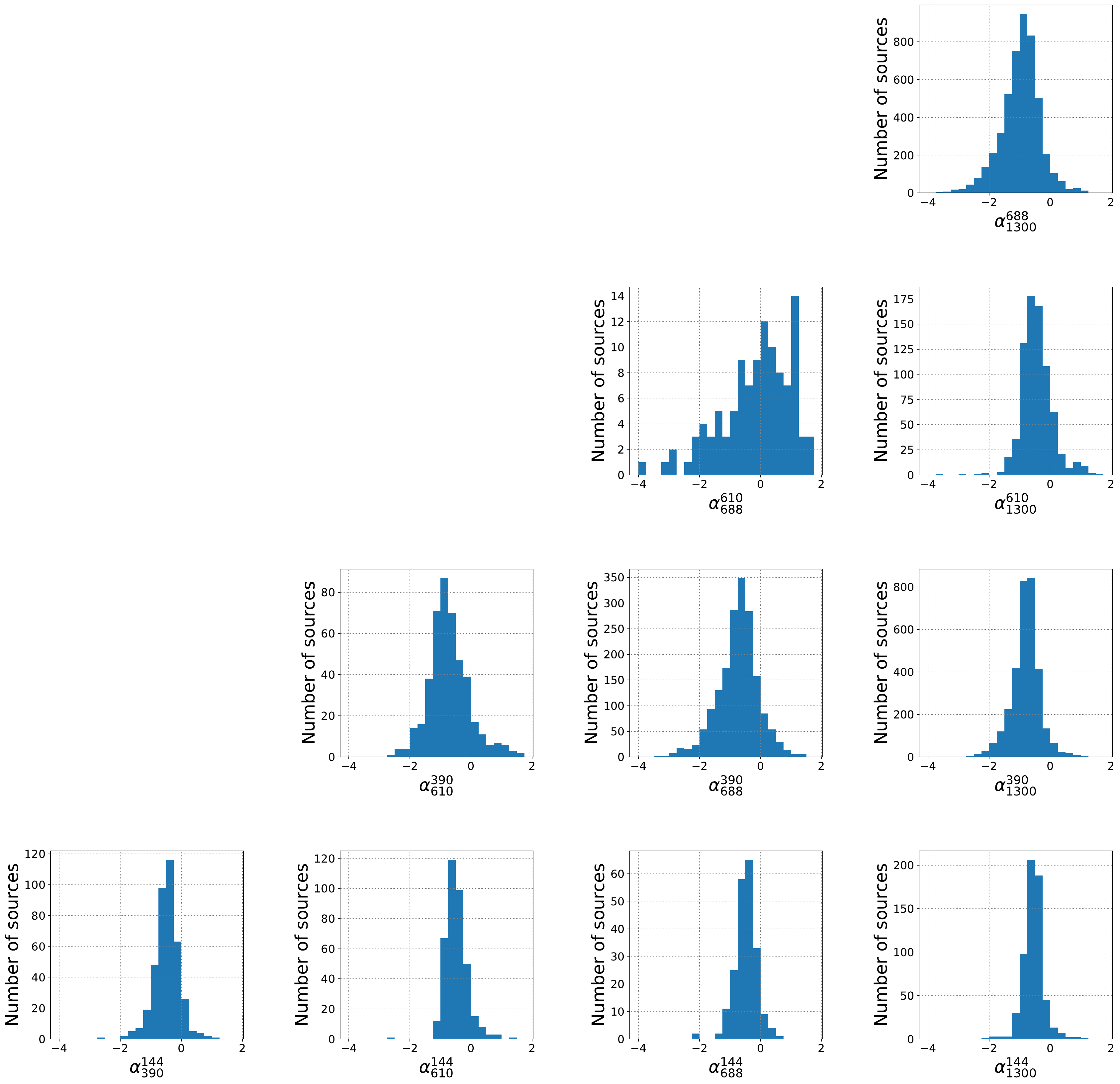}
	\caption{The distribution of the spectral index for each detected source where the spectral index is calculated for flux densities measured at 144 MHz (LOFAR), 390 MHz (uGMRT band-3), 610 MHz (GMRT), 688 MHz (uGMRT band-4), and 1.3 GHz (MIGHTEE).}
	\label{2ALdist}
\end{figure*}

\begin{figure*}
	\includegraphics[width=7.0in,height=7.0in]{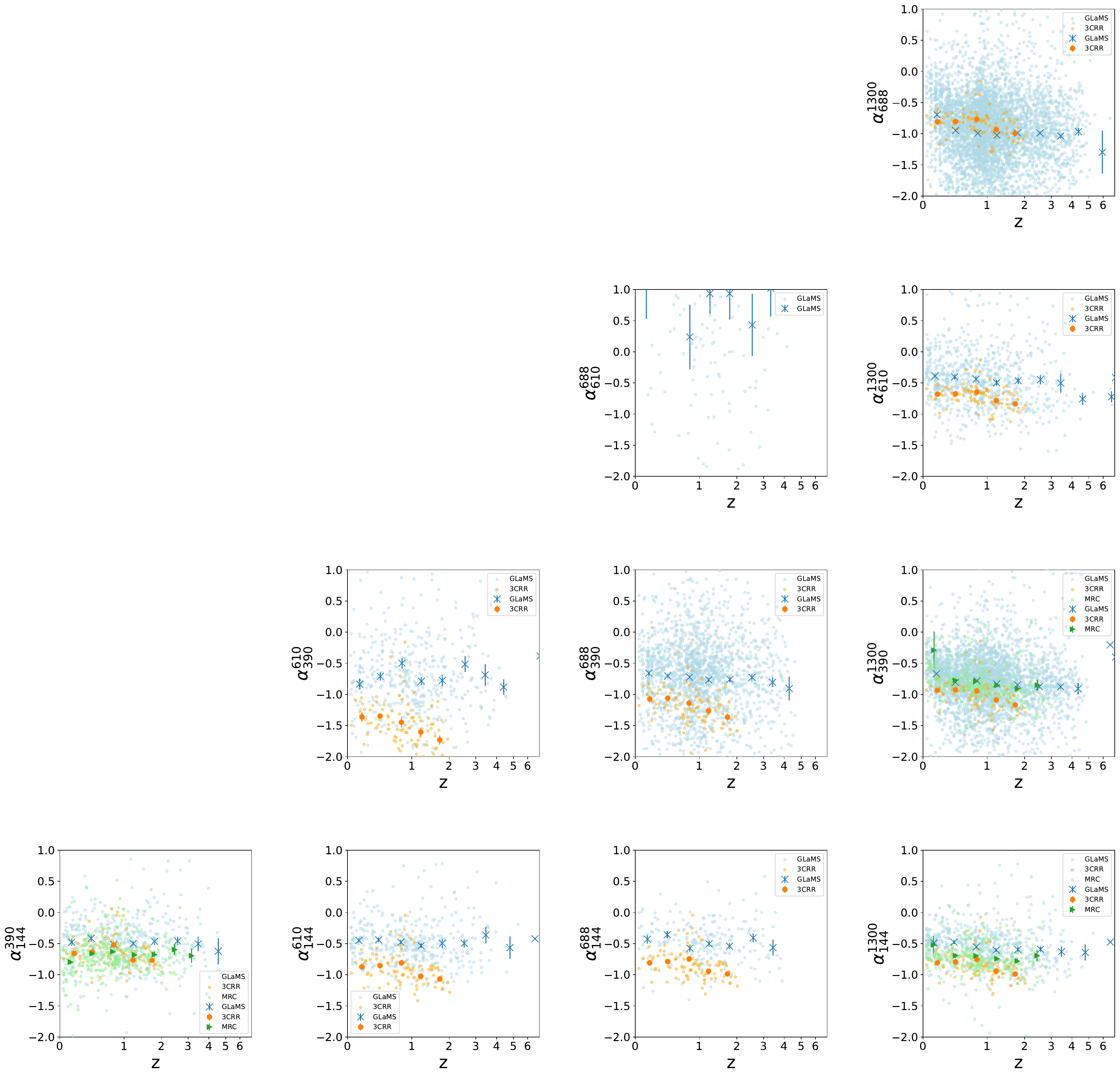}
	\caption{The correlation of the average spectral index with the redshift of the sources where the spectral index is calculated for flux densities measured at 144 MHz (LOFAR), 390 MHz (uGMRT band-3), 610 MHz (GMRT), 688 MHz (uGMRT band-4), and 1.3 GHz (MIGHTEE). Individual data points are also plotted without errors in order to indicate the spread of the data. The MRC sample is not present for some plots as the sample size for these frequencies is very low. Error bars indicate the $1\sigma$ error on the weighted mean.}
	\label{2ALZ}
\end{figure*}

\begin{figure*}
	\includegraphics[width=7.0in,height=7.0in]{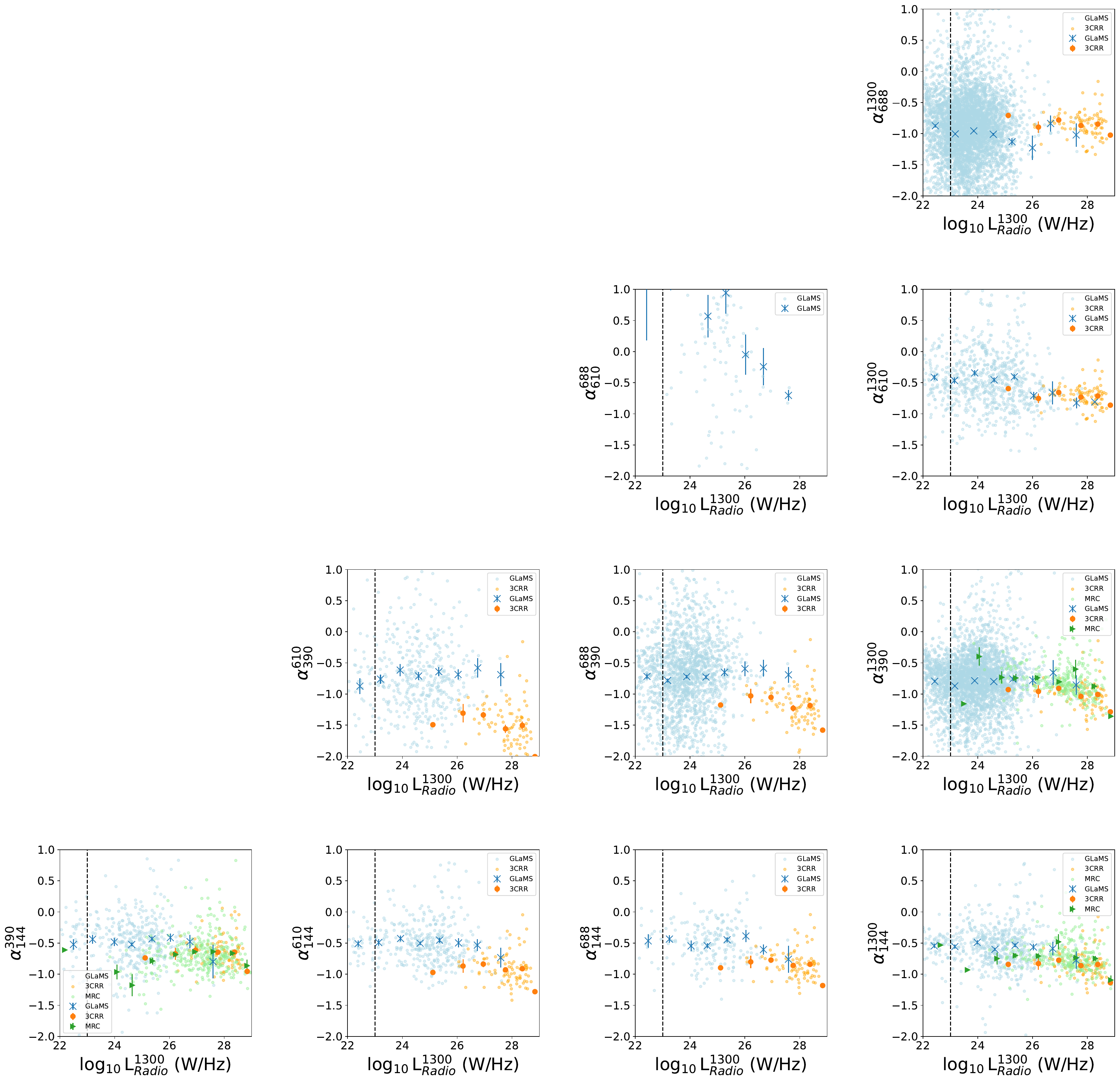}
	\caption{The correlation of the spectral index with the luminosity of the sources from MIGHTEE at 1.3 GHz where the spectral index is calculated for flux densities measured at 144 MHz (LOFAR), 390 MHz (uGMRT band-3), 610 MHz (GMRT), 688 MHz (uGMRT band-4), and 1.3 GHz (MIGHTEE). Luminosity is calculated as described in Section \ref{sec:PA}. Comments as in Fig.\ \ref{2ALZ}. The vertical dotted line shows the radio luminosity value above which AGN start to dominate.}
	\label{2ALL}
\end{figure*}

\begin{table*}
    \centering
\resizebox{\textwidth}{!}{
\begin{tabular}{c c c c c c c c c c c}
        \hline
        \\
        Survey & Frequency Range (MHz) & $N_{GLaMS}$ & $N_{3CRR}$ & $N_{MRC}$ & $\rho_{GLaMS}^{\alpha-z}$ (p) & $\rho_{3CRR}^{\alpha-z}$ (p) & $\rho_{MRC}^{\alpha-z}$ (p) & $\rho_{GLaMS}^{\alpha-L}$ (p) & $\rho_{3CRR}^{\alpha-L}$ (p) & $\rho_{MRC}^{\alpha-L}$ (p)\\
        \\
        \hline \hline
        LOFAR - MIGHTEE & 144-1300 & 602 & 90 & 288 & -0.312 (0.402) & -0.600 (0.284) & -0.428 (0.396) & -0.619 (0.101) & -0.714 (0.110) & -0.266 (0.487) \\
        LOFAR - uGMRT B4 & 144-688 & 210 & 90 & - & -0.321 (0.482) & -0.600 (0.284) & - & -0.428 (0.289) & -0.314 (0.544) & - \\
        LOFAR - GMRT & 144-610 & 378 & 90 & - & -0.016 (0.966) & -0.600 (0.284) & - & -0.476 (0.232) & -0.314 (0.544)& - \\
        LOFAR - uGMRT B3 & 144-390 & 397 & 90 & 355 & -0.428 (0.289) & -0.600 (0.284) & 0.178 (0.701) & -0.023 (0.955) & -0.028 (0.957) & 0.083 (0.831)\\
        uGMRT B3 - MIGHTEE & 390-1300 & 3219 & 90 & 309 & -0.018 (0.960) & -0.890(0.037) & -0.828 (0.041) & 0.333 (0.419) & -0.771 (0.072) & -0.350 (0.355)\\
        uGMRT B3 - uGMRT B4 & 390-688 & 1790 & 90 & - & -0.904 (0.002) & -0.890 (0.037) & - & 0.690 (0.057) & -0.771 (0.072)& - \\
        uGMRT B3 - GMRT & 390-610 & 446 & 90 & - & 0.266 (0.487) & -0.890 (0.037) & - & 0.571 (0.138) & -0.771 (0.072) & - \\
        GMRT - MIGHTEE & 610-1300 & 764 & 90 & - & -0.600 (0.066) & -0.600 (0.284) & - & -0.750 (0.019) & -0.600 (0.207) & - \\
        GMRT - uGMRT B4 & 610-688 & 159 & 90 & - & -0.357 (0.431) & - & - & -0.904 (0.002) & - & - \\
        uGMRT B4 - MIGHTEE & 688-1300 & 4851 & 90 & - & -0.616 (0.076) & -0.600 (0.284) & - & -0.357 (0.385) & -0.600 (0.207) & - \\
        \hline
    \end{tabular}}
    \caption{Number of sources in the analysis at each frequency range and correlation coefficients between spectral index, luminosity and redshift for different pairs of frequencies in the GLaMS catalogue. We use the effective frequency for the MIGHTEE sample wherever required and report the average frequency of the MIGHTEE survey in the table. The number of sources for GLaMS, 3CRR, and MRC sample are given in columns $N_{GLaMS}$, $N_{3CRR}$, and $N_{MRC}$, respectively. The correlation between spectral index and redshift is given by columns $\rho_{GLaMS}^{\alpha-z}$ ($p$), $\rho_{3CRR}^{\alpha-z}$ ($p$), and $\rho_{MRC}^{\alpha-z}$ ($p$). The correlation between spectral index and radio luminosity is given by columns $\rho_{GLaMS}^{\alpha-L}$ ($p$), $\rho_{3CRR}^{\alpha-L}$ ($p$), and $\rho_{MRC}^{\alpha-L}$ ($p$). The values in the brackets give the $p$-value of the correlation, where we take a correlation with $p<0.05$ to be statistically significant. We only show MRC sources where the number of sources in the MRC sample is greater than 20.}
    \label{Table:2pSampleSize}
\end{table*}

In this section we show and discuss the various combinations of two point analysis at different frequencies and frequency ranges. We show the overall distribution of the spectral index for the GLaMS sample for pairs of frequency ranges in Fig. \ref{2ALdist}. In Fig. \ref{2ALZ}, we present spectral index versus redshift for the different values of spectral index obtained by using various combinations of frequencies, while the number of sources in the GLaMS sample are given in Table \ref{Table:2pSampleSize}. The errors on the spectral index are obtained using error propagation of flux density errors for each source in the sample. We bin all the detected sources for a given frequency pair in redshift and evaluate the mean spectral index values, where uncertainties are calculated using the standard error on the mean. In Table \ref{Table:2pSampleSize}, we also report the Spearman rank correlation along with their $p$-values for each frequency pair. These correlation values are obtained by using the mean binned values. 

We compare results for various frequency ranges, which include both results for the smallest frequency difference such as GMRT 610-MHz to uGMRT band-4 at 688 MHz, and results from the largest frequency difference, i.e. LOFAR 144 MHz to MIGHTEE 1300 MHz. We repeat the same analysis for the sample obtained from the 3CRR survey and the MRC survey to present a comparison with the data from previous studies. The plots in Fig. \ref{2ALZ} show the mean of the spectral index as a function of redshift. Frequency range pairs such as uGMRT-B4 - MIGHTEE, uGMRT-B3 - MIGHTEE, and uGMRT-B3 - B4 have the highest number of sources as the frequency ranges are from the sensitive uGMRT and MIGHTEE survey whereas pairs such as LOFAR - uGMRT-B3 and LOFAR - GMRT both contain data from the less sensitive LOFAR survey. In all of the plots of Fig. \ref{2ALZ} we can see that the spectral index values for GLaMS are comparatively flatter than the spectral index values observed for the 3CRR survey and the MRC spectral index values lie in between the two, especially for the frequency ranges with 200 or more objects. We also observe an offset between the $\alpha$ values of GLaMS, 3CRR, and MRC sample, for the same redshift bins. We note from Table \ref{Table:2pSampleSize} that the GLaMS sample shows lower correlation values between spectral index and redshift than 3CRR, except for the uGMRT B3-B4, GMRT-MIGHTEE, and uGMRT B4-MIGHTEE correlations, although uGMRT B3-B4 and GMRT-MIGHTEE still show an offset between GLaMS and 3CRR spectral index values.\footnote{The GMRT to uGMRT-B4 pair in Fig. \ref{2ALZ} is shown for completeness, but is not reliable because of the very close frequencies for these objects.}

We also note that the $p$-values for the $\alpha$-$z$ relation in GLaMS exceed the 5\% threshold for all frequency pairs -- there is no statistically significant $\alpha$-$z$ correlation in the binned GLaMS data.  This is not true for the 3CRR sample where at least three frequency pairs, corresponding to uGMRT B3 - MIGHTEE uGMRT B3 - uGMRT B4, and uGMRT B3 - GMRT, show $p$-values less than 5\% implying a significant correlation. The MRC sources behave consistently with 3CRR, at least at the frequencies where we have sufficient data to make the comparison. 
   
In order to understand the discrepancy observed in Fig. \ref{2ALZ}, between the GLaMS sample and the 3CRR sample, we look at the luminosities of the samples, as we know that the selections made in the older samples are dominated by luminous sources. Fig. \ref{2ALL} shows plots of spectral index and the radio luminosity of the sources in the GLaMS sample, the 3CRR sample, and the MRC sample. The spectral index values for sources are averaged for a radio luminosity bin size of 0.5 decades of radio luminosity for both sets of observations. From all the plots in Fig. \ref{2ALL}, we can see that all the 3CRR sample contains only luminous sources, i.e. higher than L$_{1300}$ $10^{25}$ W Hz$^{-1}$. In addition, we also see for all plots that the 3CRR sample shows a downward trend, which indicates that as the luminosity increases the spectral index steepens. Further, when we look at the GLaMS sample and its arrangement in the plots we can see that the trend from the 3CRR sample is continued to lower luminosities by the GLaMS sources, following a similar slope to that exhibited by the 3CRR sample for most plots, except for uGMRT B4 - B3 and GMRT 610 - B3, where the two samples are offset from each other. From Table \ref{Table:2pSampleSize}, we observe that the correlation values in the GLaMS sample are mostly closer to the values found for the 3CRR samples, for examples, LOFAR - MIGHTEE, LOFAR - uGMRT B4, LOFAR - GMRT, GMRT - MIGHTEE, and the uGMRT B4 - MIGHTEE. Out of these, GMRT - MIGHTEE, and GMRT - uGMRT B4 show significant correlations with $p<0.05$. We also observe that the uGMRT B3-B4, uGMRT B3-MIGHTEE, and uGMRT B3-GMRT pairs shows a positive correlation with luminosity, although looking at the plot in Fig. \ref{2ALL}, we can see that the mean spectral indices are almost constant with luminosity for the frequency pairs (noting that the sample size for uGMRT B3-GMRT is low and the data show high scatter). In addition, the plots of the three pairs show lower number of sources at higher radio luminosity bins for GLaMS, which could shift the average spectral index to flatter values and be responsible for overall positive correlation. Overall, both from the individual data points and the mean values, we see that in most cases the trend in the GLaMS sample is continued by the 3CRR sample and almost all of them show a slight downward progression in the sense that higher luminosities imply steeper spectra. The same is also evident from the background scatter shown in the plots. The vertical dotted line shown in the plots represents the boundary before which the dominance of SFGs is prominent. From the plots, only one or two data points fall within the limit and hence the presence of SFGs in our sample does not affect our overall result. 

The results of studies such as those of \cite{Gopalkrishnaetal1988}, \cite{Onuoraetal1989}, and \cite{Blundelletal1999}, also suggest a relationship between luminosity and spectral index, but our work extends the luminosity dependence to even lower luminosities and to much larger sample sizes. In our analysis we have observed a statistically significant luminosity-spectral index correlation for some frequency pairs, although the relationship is not very prominent until we look at the background scatter. \cite{Blundelletal1999} proposed that the spectral index/luminosity relation effect can give rise to an apparent spectral index/redshift relation, as luminosity is a function of spectral index and redshift for flux limited samples. For a given source at high redshift, the luminosity of the source needs to be high enough to be detected in the survey, depending on the instrument's flux limits. As we cannot observe low-luminosity sources below some threshold at higher redshifts, the spectral index to redshift correlation becomes more evident in such cases, which shows up for surveys like 3CRR but is not seen in the GLaMS sample where low-luminosity sources are seen at all redshifts. However, there are other effects, such as selection of sources, that can mask this one: in this paper we have only selected compact sources and effects from extended sources have not been included in the analysis.

\begin{figure*}
	\includegraphics[width=3.5in]{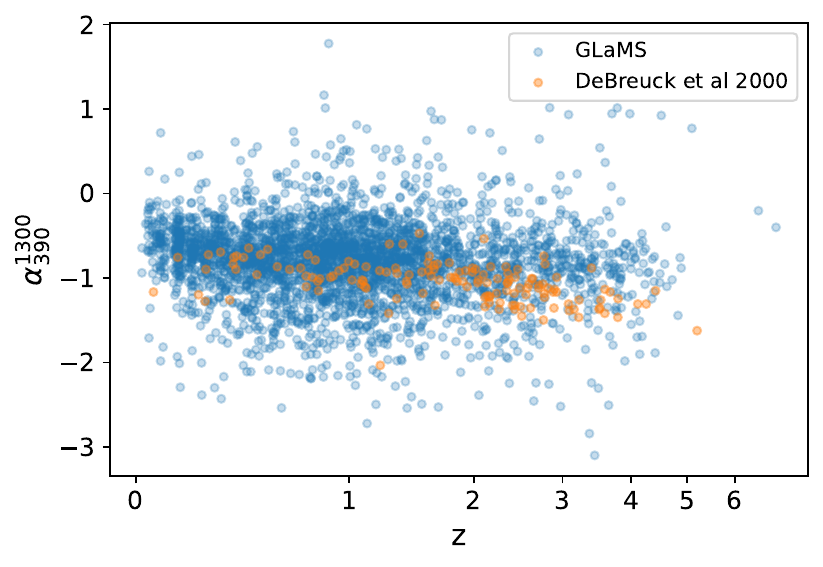}\includegraphics[width=3.5in]{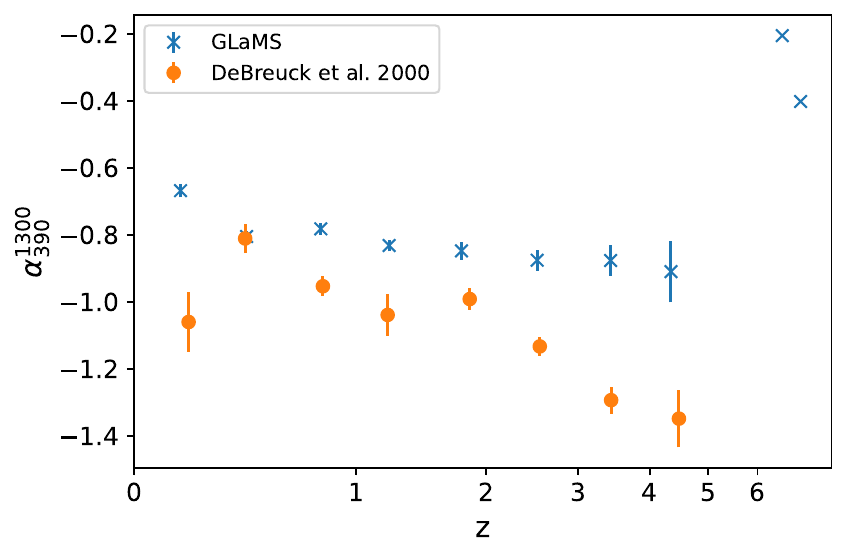}
    \includegraphics[width=3.5in]{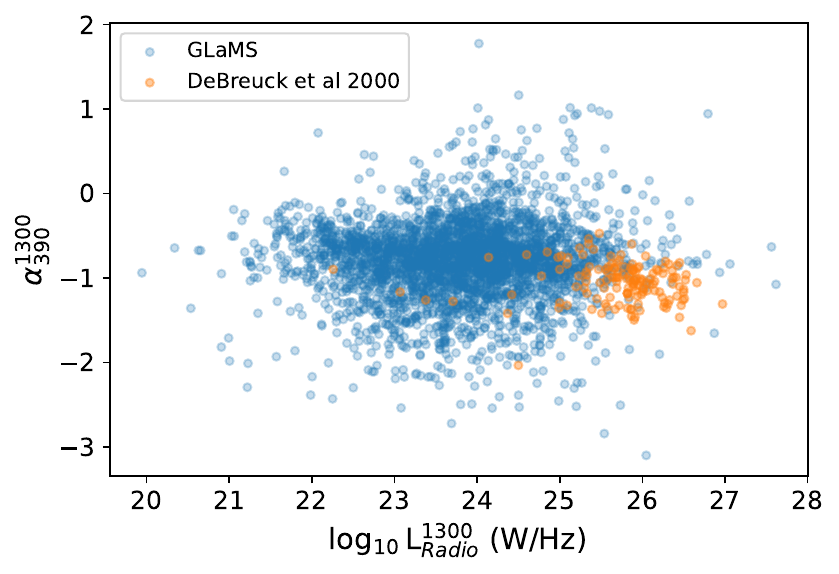}\includegraphics[width=3.5in]{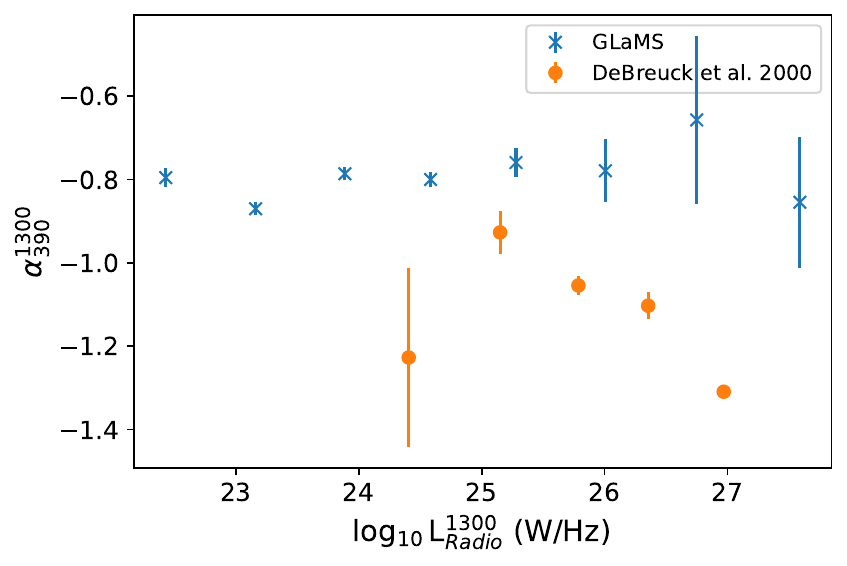}
	\caption{Top row: the relationship  between the spectral index and the redshift, comparing the observations of sources obtained by \protect\cite{Debrucketal2000a} and the GLaMS sample: on the left we show the distribution of individual sources and on the right the average values in matched bins in redshift are shown. Bottom row: the same relationship between radio luminosity and spectral index, with a radio luminosity bin size of 1 decade in radio luminosity for both the samples.}
	\label{Debruck_glams}
\end{figure*}  

The study of steep-spectrum radio sources by \cite{Debrucketal2000a} is widely cited as showing a strong spectral index/redshift correlation, with an almost linearly increasing correlation between the steepness of the spectrum and the redshift for their sample. They present flux densities and spectral indices for 147 sources at frequencies of 325 and 1400 MHz, very close to our uGMRT B3 and MIGHTEE observing frequencies respectively, which allows a direct comparison. As shown in Fig. \ref{Debruck_glams} (top row), we do observe the same for the GLaMS sample over the same frequency range, although the trend is comparatively flatter in the GLaMS sample and there is an offset between the spectral index values for a given redshift. As above, we note that one of the major differences between the GLaMS sample and the \cite{Debrucketal2000a} sample is that the GLaMS sample consists of low luminosity sources as compared to those studied by \cite{Debrucketal2000a}, also shown in Fig. \ref{Debruck_glams}, in the bottom row. We see from the same figures that the slope of either the luminosity or redshift correlations from the GLaMS sample is flatter than that of the \cite{Debrucketal2000a} sample. This reinforces the point already seen from the 3CRR and MRC sources above: there is a trend for spectra to be steeper at higher redshift in both the GLaMS sample and the comparison samples, but they are not the same trend.   

Finally, we considered the possibility that the systematic offset in the $\alpha$-$z$ relations between the GLaMS objects and earlier samples might be due to the fact that the latter include extended sources whereas our study does not. We cannot include extended sources in our sample but we can test the effects of excluding them in the case of the 3CRR sample, where largest angular size measurements are available for all sources. When this test is carried out we see no clear difference between the trends for the small-source subset of 3CRR and the whole sample, and the $\alpha$-$z$ offset is still clearly visible. Full details of this test and its results are presented in Appendix \ref{app:3csubset}.

\subsection{Multi-frequency analysis}
\label{3FA}
\begin{figure*}
	\includegraphics[width=3.5in]{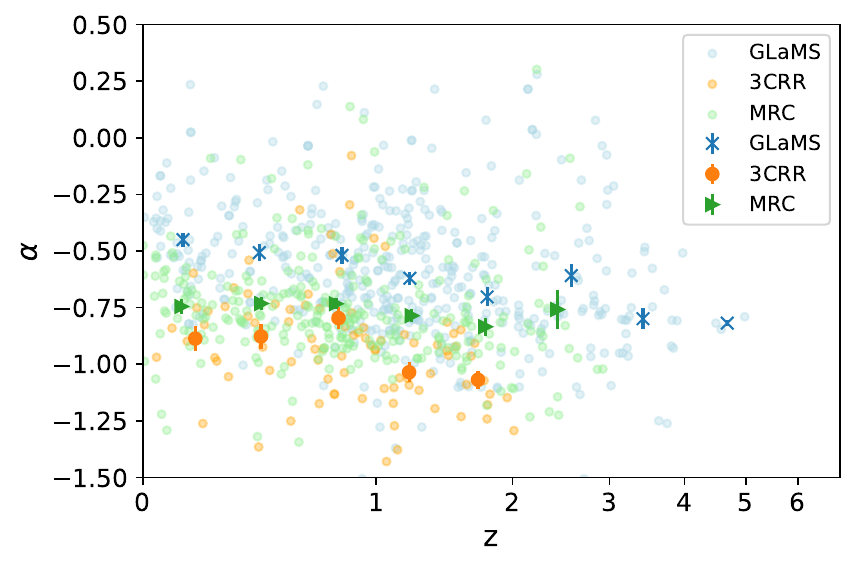}\includegraphics[width=3.5in]{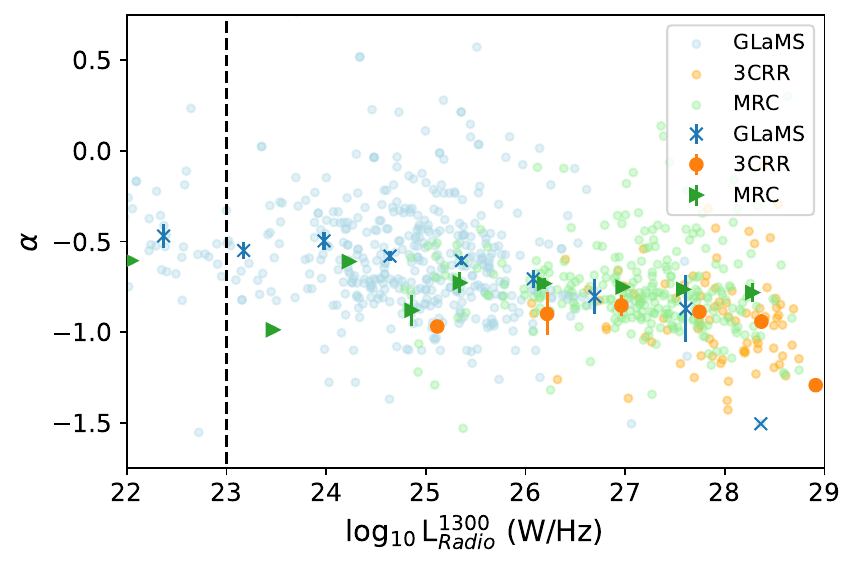}
	\caption{Left: correlation between the spectral index and the redshift (left) obtained by evaluating spectral index using flux from three or more frequencies. Right: correlation between the radio luminosity and the spectral index (right) for the same data. The vertical dotted line shows the radio luminosity value after which AGN start to dominate.}
	\label{allALZ}
\end{figure*}

\begin{table}
    \centering
\begin{tabular}{c c c c}
        \hline
        \\
        Sample & Number of Sources & $\rho^{\alpha-z}$ (p) & $\rho^{\alpha-L}$ (p)\\
        \\
        \hline \hline
        GLaMS & 522 & -0.683 (0.042) & -0.809 (0.014) \\
        3CRR & 90 & -0.600 (0.284) & -0.257 (0.622) \\
        MRC & 287 & -0.657 (0.156) & -0.283 (0.460) \\
        \hline
    \end{tabular}
    \caption{Correlation analysis for each sample for the spectral index, redshift, and luminosity. We use effective frequency for the MIGHTEE sample and the number of sources for GLaMS, 3CRR, and MRC sample are given in column 'Number of Sources', respectively. The correlation between spectral index and redshift is given by column $\rho^{\alpha-z}$ (p). The correlation between spectral index and radio luminosity is given by column $\rho^{\alpha-L}$ (p). The values in the bracket give the p-value of the correlation.}
    \label{Table:MultiSample}
\end{table}

In this section we explore the multi-frequency spectral index relation with the redshift and the radio luminosity, where Fig.\ref{allALZ} shows the plots of spectral index versus redshift on the left and spectral index versus radio luminosity on the right. As stated in Section \ref{Intro}, using flux from multiple frequencies as input points we can fit a power law to all the flux measurements to get the spectral index based on multiple data points. We do this using the {\sc scipy} function \citep{SciPy} {\tt curve\_fit} in linear space taking account of the error bars, i.e. $\chi^2$ minimization. We force the fits to include the LOFAR survey and the MIGHTEE survey by selecting only sources that have data at the respective frequencies. We do this because they represent the largest frequency range we have for our sample and the MIGHTEE survey is the most sensitive survey we have in our sample.  We also select sources that also have flux values present for at least one of the other three surveys, i.e. the GMRT survey and the uGMRT band 3 and band 4 survey. By selecting sources with flux values at three or more frequencies we obtain spectral index values obtained from multiple bands, which allows us to improve the accuracy of the broad-band alpha values \footnote{We explicitly chose not to determine a spectral index for all sources for which three, or even two frequencies were available, but to require the frequency coverage to span the 144-1300 MHz range given by LOFAR and MIGHTEE. This is because any given pair of frequencies suffers from bias as shown in Fig.\ \ref{FB}; a sample constructed using all available pairs of frequencies would have a {\it flux-dependent} bias and that could result in spurious correlations in the $\alpha$-$z$ or $\alpha$-$L$ plots.}. The number of sources obtained using this method for the three samples is given in Table \ref{Table:MultiSample}; the number of GLaMS sources is significantly reduced by the requirement to include LOFAR data in the analysis. We use the 90 sources from the 3CRR survey as we have values at all the frequencies. For MRC we do the same by selecting sources at the respective frequency range used in the 3CRR sample where we ignore the frequency band between 600 MHz and 800 MHz as requiring these would reduce the sample size to less than 10. The spectral index values shown in the plots of Fig.\ref{allALZ} are averaged for a redshift bin size of 0.1 in $\log_{10}(1+z)$ for the left panel and a radio luminosity bin size of 0.5 decades in $\log_{10}(L_{1300})$ for the right panel. In Table \ref{Table:MultiSample}, we also report the Spearman correlation for the spectral index versus redshift and spectral index versus luminosity plots.  

From Fig.\ref{allALZ}, we find a similar trend between the spectral index and redshift to the one that we observed in the two point frequency analysis. We can see that the GLaMS data show evolution with $z$, although we again observe an offset between the 3CRR and GLaMS spectral index values with the MRC objects lying between the two. For the 3CRR sample, the spectral index tends to steepen with increasing redshift and the steepening is comparatively stronger than for GLaMS or MRC. For the MRC sample, we see a trend similar to the 3CRR sample at least for the first few data points after which the scatter increases. This is also representative of the correlation values observed in Table \ref{Table:MultiSample}. The spectral index of the MRC sample is also intermediate between the spectral index of the GLaMS and the 3CRR sample which suggests a dependency of these apparent trends on the luminosity. To confirm this we can look at the right plot of Fig. \ref{allALZ} and the correlation values in Table \ref{Table:MultiSample}, where we can see clearly that the data points from the three samples lie on the same trend of steeper spectrum for higher luminosity. This is a clear indication that a relationship between luminosity and spectral index exists, as discussed in the previous section.    

\subsection{Interpretation of the correlations}
\begin{figure*}
	\includegraphics[width=3.8in,height=2.8in]{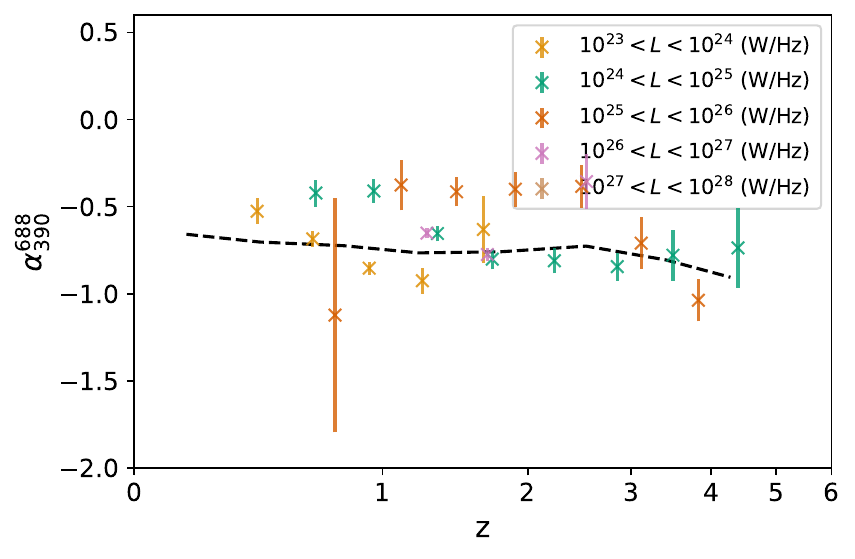}\includegraphics[width=3.8in,height=2.8in]{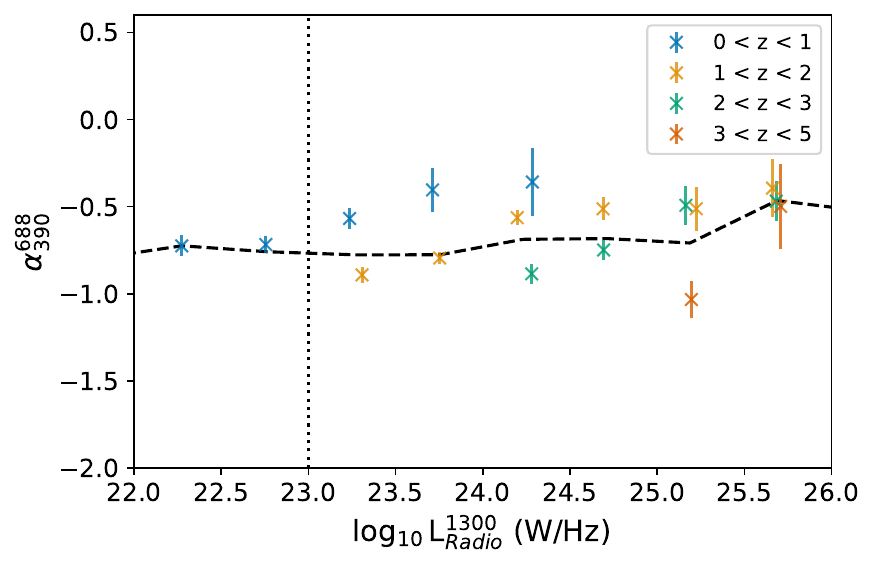}
    \includegraphics[width=3.8in,height=2.8in]{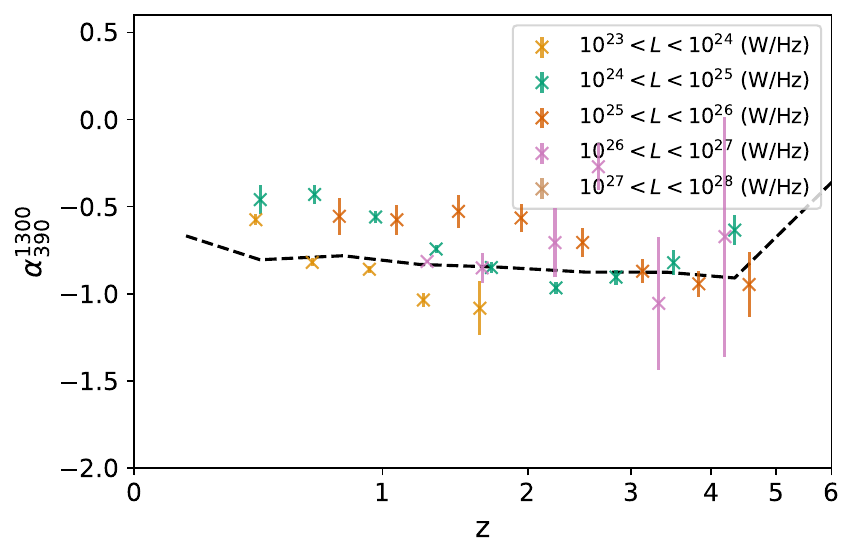}\includegraphics[width=3.8in,height=2.8in]{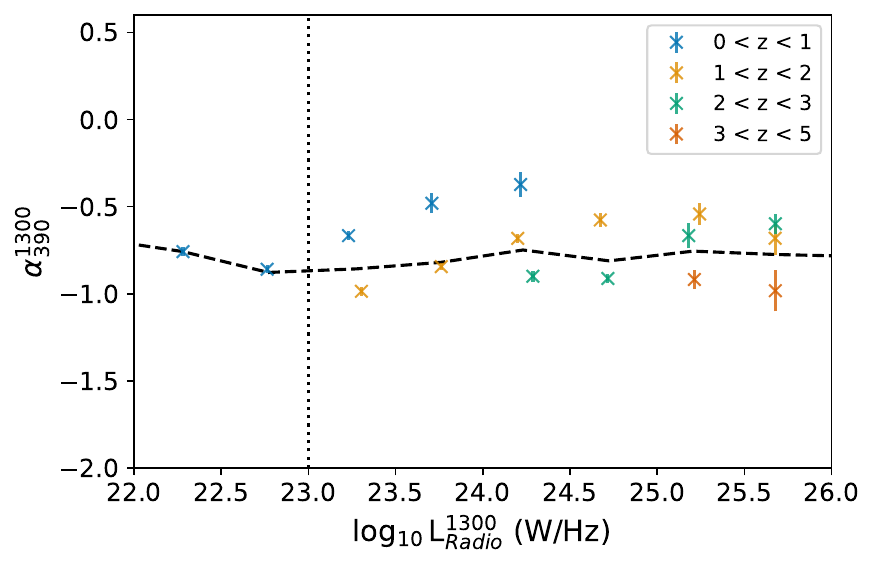}
    \includegraphics[width=3.8in,height=2.8in]{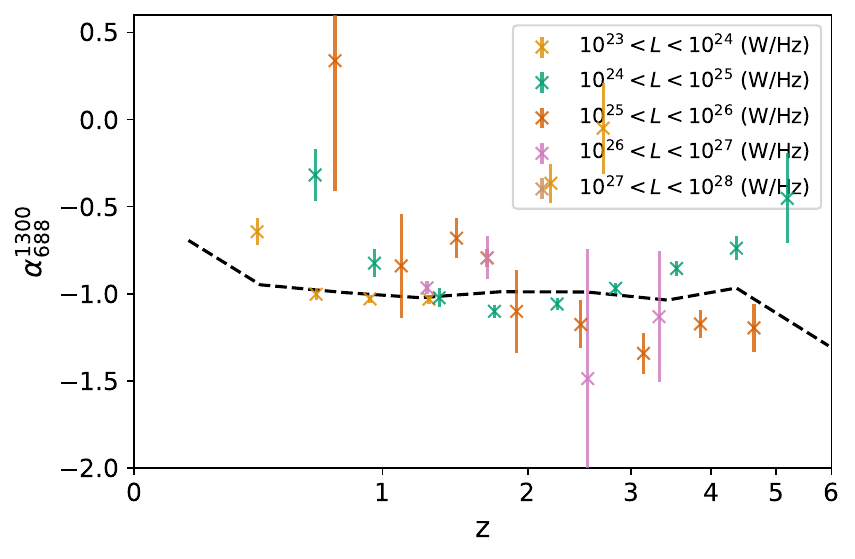}\includegraphics[width=3.8in,height=2.8in]{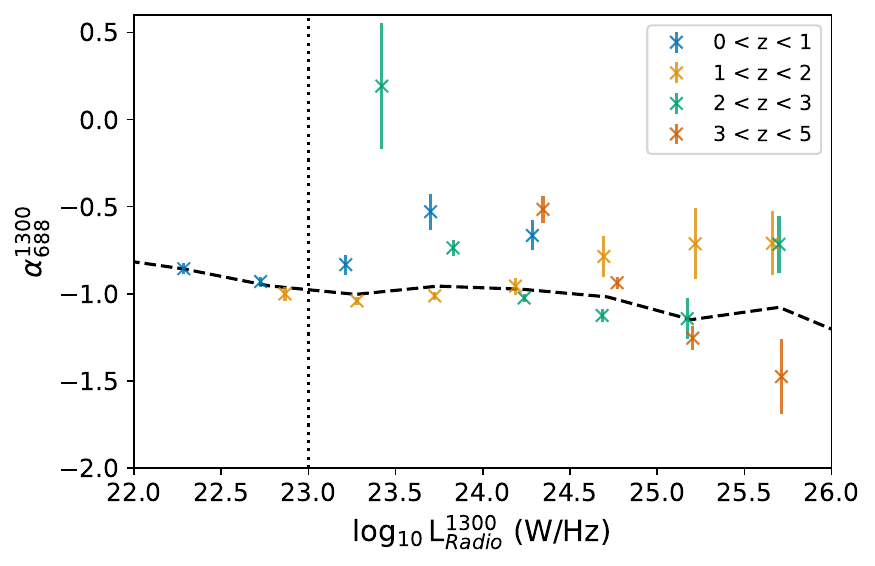}
	\caption{Left column: the relationship  between the two-point spectral index and the redshift, filtered for different range of radio luminosities. Right column: relationship between spectral index and radio luminosity, filtered for different ranges of redshifts. The black dashed line represents the mean of $\alpha$ as seen for the frequency pairs in Fig \ref{2ALZ} and \ref{2ALL}, averaging over all GLaMS sources at these frequencies. The vertical dotted line shows the radio luminosity value above which AGN start to dominate.}
	\label{Alpha_L_z_filters}
\end{figure*}  

\begin{figure*}
	\includegraphics{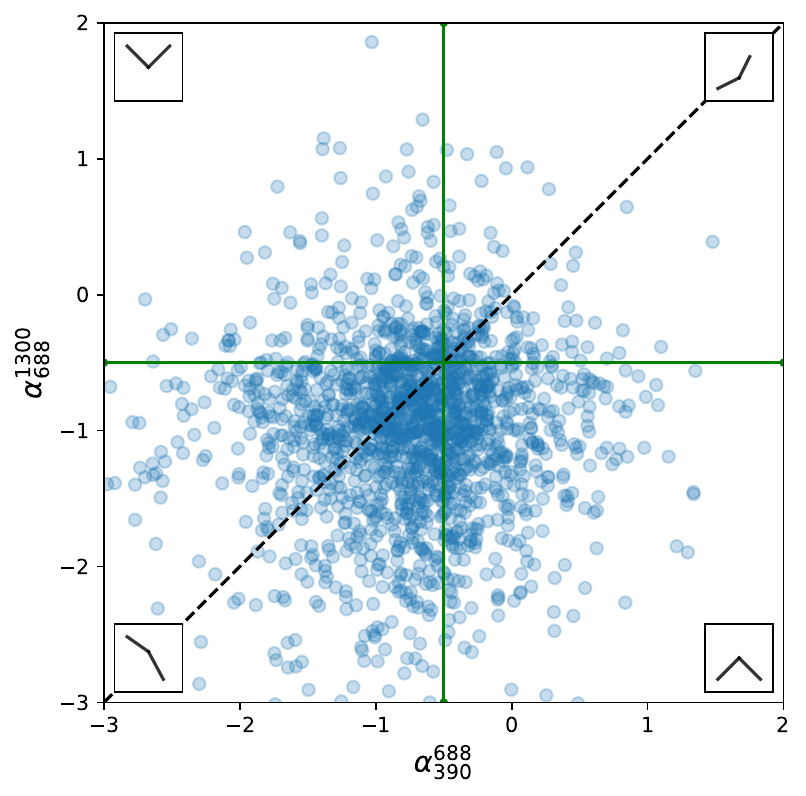}
	\caption{Plot showing correlation between the spectral index obtained for frequency range uGMRT B3 - B4 and uGMRT B4 - MIGHTEE, divided into quadrants at a spectral index of $-0.5$ which is the flattest value expected for optically thin synchrotron emission. The four small plots on each quadrant illustrate the respective spectral curve. The dashed line shows the line of equality of spectral index at the two frequencies.}
	\label{AlphaB3B4_B4M}
\end{figure*}  
From both the preceding subsections we have seen weak but significant correlations between redshift and spectral index, and luminosity and spectral index, for the GLaMS sources. However, the GLaMS sources show systematically flatter spectral indices than 3CRR or MRC sources at the same redshift, while plots of spectral index against luminosity show broadly similar trends of steeper spectra with higher luminosity in all three samples. We interpret the fact that all the samples seem to lie on the same spectral index/luminosity trend in terms of a direct relationship between luminosity and spectral index.

What could cause this relationship? \cite{Blundelletal1999} pointed out that a relationship between luminosity and spectral index that persists over a wide range of frequencies is most easily explained in terms of the injection index, i.e. the spectral index of particles when they are originally detected. The most powerful radio AGN are FRII sources where the main location of particle acceleration is the hotspots, and in these we expect higher jet powers to equate to higher magnetic field strengths and synchrotron photon densities, and thus to higher radiative and inverse-Compton losses, so that qualitatively the energy spectrum of particles escaping from the hotspot might be expected to steepen with increasing jet power, while potentially still being flatter at the lowest energies. They argued that the non-detection of optical synchrotron emission from the most powerful hotspots is evidence that the synchrotron spectra may be steeper in those systems \citep[cf.][]{Meisenheimeretal1997, Brunetti2003}. Since then the widespread detection of X-ray synchrotron from lower-power hotspots \citep{Hardcastleetal2004} has provided evidence that the overall synchrotron spectrum, including the high-energy cutoff, depends on jet power in some way. Further evidence supporting this general picture comes from the observation that the pairs of hotspots in double-double radio galaxies have the same spectral indices, despite their very different dynamics \citep{Konaretal2003}, which can only be explained in terms of a direct jet power/hotspot spectrum relationship.


In an effort to explore the spectral index, redshift, and luminosity relationship for the GLaMS sources in more detail we have generated $\alpha$-$z$ and $\alpha$-$L$ plots for the frequency pairs that have the largest sample sizes and are most sensitive, that is, the uGMRT B3 - B4, uGMRT B4 - MIGHTEE, and the uGMRT B3 - MIGHTEE pairs. For these plots we filter samples in different redshift ranges for the $\alpha$-$L$ plots and filter samples in different radio luminosity ranges for the $\alpha$-$z$ plots in order to separate out the redshift and luminosity effects. The plots are shown in Fig. \ref{Alpha_L_z_filters}. These plots show that the picture is more complicated than is consistent with a simple luminosity/spectral index relationship. In fixed luminosity bins (left column) we see a steepening spectral index as a function of redshift in almost every bin, though this is more prominent at the two lower frequencies, and is only modest in magnitude (e.g. sources in the bin $10^{25}$ to $10^{26}$ W Hz$^{-1}$ have typical spectral indices that steepen from $-0.5$ to $-0.9$ between $z=1$ and $z=4$). Moreover, in a given redshift range (right column), more luminous sources tend to have \textit{flatter} spectra, which is the opposite of what would be predicted by the \cite{Blundelletal1999} model or what is expected from the offset between 3CRR and GLaMS sources.

The steepening of spectral index with redshift and luminosity is consistent with the expected effect of inverse-Compton losses and higher radiative losses, combined with the fact that we observe at higher rest-frame frequencies at higher redshifts\footnote{In our sample there may also be more subtle effects such as a bias against physically large sources at low redshift due to our angular size cutoff; exploring the effects of these will have to await the availability of a full optical identification for the larger MIGHTEE sources.}. However, we are not sure why low luminosity sources are steeper-spectrum than high luminosity sources, especially for the uGMRT B3 - B4 and uGMRT B3 - MIGHTEE pairs. Most of the sources driving these trends are much lower in luminosity than the sources discussed by \cite{Blundelletal1999}, and it may be that particle acceleration operates differently in these low-luminosity objects, or that they have a larger fraction of sources affected by self-absorption or free-free absorption. At the lowest luminosities, many may not be AGN at all. In Fig. \ref{AlphaB3B4_B4M}, we show a scatter plot of the spectral indices for the two frequency ranges discussed above. We can see that there are around 542 sources in the quadrant where the spectrum turns down at low frequencies, consistent with the idea that one or both of the absorption processes are important for a significant fraction of our sample. To take this analysis further it will also be important to consider the full population of MIGHTEE sources by including the extended objects when they have all been identified, to include a full coverage of physical sizes in our sample, although as discussed above we have reason to believe that our results are not driven by the missing extended sources in GLaMS.

Finally, we checked the robustness of our results by conducting the same analysis using only sources that have flux density greater than $10^{-4}$ and $10^{-3.5}$ Jy for uGMRT B4 and B3 respectively. As we can see from Fig.\ref{FB}, at these flux limits the sensitivity limit intersects the spectral index line at a spectral index of 1 which essentially removes any bias against inverted-spectrum sources. We find qualitatively similar results after conducting this analysis and conclude that the bias against inverted spectrum sources does not have a significant effect on our results.

\section{Conclusions}
\label{Con}
We have used the data from five different surveys carried out in the XMM-LSS field to look at the spectral index behavior of radio sources as a function of redshift and luminosity. We used two point spectral index analysis and multi-frequency spectral index analysis to revisit the correlation between spectral index, luminosity and redshift for much larger samples than have hitherto been available and over a wide range of different combinations of frequency. As investigated by different studies, discussed in Section \ref{Intro}, it has been observed that there is a positive correlation between the spectral index and the redshift, i.e. the spectral index of the sources become steeper with increasing redshift. In the past, this correlation has been used to identify steep spectrum sources, especially for high redshift radio galaxies. However, the correlation has largely been explored for bright sources such as those from the 3CRR and MRC surveys \citep{Debrucketal2000a, Morabitoetal2018}. 

From the results obtained using the two-point analysis and multi-frequency analysis for the three samples, we can answer the questions presented in the Section \ref{Intro}, which are as follows:
\begin{enumerate}[i]
	\item We observe that for our sample the spectral index increases weakly but significantly in many cases with redshift. However, we also observe an offset between the mean spectral index values obtained from GLaMS and 3CRR: at the same redshift, the more luminous 3CRR sources show systematically steeper spectra. 
	\item We observe a weak but again significant correlation between the radio luminosity and averaged spectral index for most frequency pairs, although some pairs, such as LOFAR - MIGHTEE and GMRT - MIGHTEE, show a more prominent increasing trend. The more luminous 3CRR and MRC sources that we compare with lie on the same trend. 
	\item In the two-point analysis we constructed ten different plots using different combinations of the frequency ranges obtained from the surveys for the two correlations. For all the plots we can see a correlation for the GLaMS sources but a stronger and more rapidly increasing trend for the 3CRR sample for the spectral index versus the redshift. By contrast, the trend between the luminosity and the spectral index is consistent with same continued slopes for the three samples in most of the pairs of frequencies we used. Due to the low number of sources in the MRC sample we observe a significant scatter in the plots. Very similar results are obtained in the multi-frequency analysis.  
	\item For the two-point analysis, the largest sample size is 4851 sources and the smallest sample size is 159 sources although the trend observed for these are more tightly constrained for almost all the large sample plots in GLaMS. The smaller samples lead to large uncertainties when binned by luminosity or redshift. A sample size of more than 500 sources is ideal to analyse such correlations if sensitive data are used.
    \item Attempting to disentangle the redshift and luminosity relations in the GLaMS sample, we find evidence for relationships between spectral index and both redshift and luminosity: in fixed luminosity bins there is a clear redshift dependence at some frequencies, while in fixed redshift bins there is a luminosity dependence.
    \item As argued by \cite{Blundelletal1999}, the relationship between spectral index and luminosity seen in luminous sources could be due to the injection index of the sources, where high power sources have high jet energy densities with stronger magnetic fields, leading to higher synchrotron losses and hence steeper spectra. This would be consistent with the very clear offset between the spectral indices of GLaMS and 3CRR sources at a fixed redshift and the continuity of the spectral index values between GLaMS and 3CRR at high luminosities. However, there is clearly also evidence for a direct relationship between spectral index and redshift at a fixed luminosity, which could be explained qualitatively in terms of increased inverse-Compton losses together with the higher rest-frame frequency of observation. The fact that spectral index in some bands shows a positive correlation with luminosity (higher luminosity gives flatter spectrum) at fixed redshift is a puzzle in this scenario, but is probably driven by the presence of many very low-luminosity sources in our sample. We would not necessarily expect the low-luminosity sources, which will be of FRI-type, to obey the same relation as the sources discussed by \cite{Blundelletal1999}, which are all powerful FRII sources with hotspots.
\end{enumerate}       


Further investigations are required, where we can explore samples from more sensitive surveys and also include extended sources in the analysis to form a complete sample, but our basic conclusion is that at high luminosities the radio luminosity is the driver of the observed steep spectra, with any direct correlation with redshift being a weaker effect. Thus we predict that ultra-steep-spectrum selection will become less and less effective to select high-redshift sources as it is applied to fainter sources with intrinsically lower luminosities.

Other studies, such as those of \cite{Anfangetal2021} and \cite{Anfangetal2024}, suggest no strong or obvious correlations between radio spectral index and redshift. These studies found results using a sample that contains SFGs and argue that including AGN does not affect their statistical results on the radio spectral index. Large sample sizes ($> 1000$) are important to analyze such correlations, as data from sensitive surveys gives larger samples and gives rise to less noisy plots. We have observed that the quality of the relationship is also improved by making use of as broad a frequency range as possible. For the XMM-LSS survey, more sensitive data at low frequencies could further help to reduce the scatter and increase the number of sources. Further investigation of these correlations could also be carried out using the LoTSS wide-area survey of the northern sky \citep{Shimwelletal2022}, where spectral index measurements are in principle available for large numbers of optically identified sources~\citep{Hardcastleetal2023}. Extension of the MIGHTEE survey to a wider range of frequencies would also allow us to expand the scope of this work. 

\section*{Acknowledgments}
MJH acknowledges support from the UK STFC [ST/V000624/1]. DVL acknowledges the support of the Department of Atomic Energy, Government of India, under project no. 12-R\&D-TFR-5.02-0700. DJBS acknowledges support from the UK Science and Technology Facilities Council (STFC) via grant numbers ST/V000624/1 and ST/Y001028/1. MJJ acknowledges support from the STFC consolidated grants [ST/S000488/1] and [ST/W000903/1] and from a UKRI Frontiers Research Grant [EP/X026639/1], which was selected by the ERC. MJJ, CLH and IHW acknowledge support from the Oxford Hintze Centre for Astrophysical Surveys which is funded through generous support from the Hintze Family Charitable Foundation. MV acknowledges financial support from the Inter-University Institute for Data Intensive Astronomy (IDIA), a partnership of the University of Cape Town, the University of Pretoria and the University of the Western Cape, and from the South African Department of Science and Innovation's National Research Foundation under the ISARP RADIOMAP Joint Research Scheme (DSI-NRF Grant Number 150551) and the CPRR HIPPO Project (DSI-NRF Grant Number SRUG22031677). JA and DB acknowledge financial support from the Science and Technology Foundation (FCT, Portugal) through research grants PTDC/FIS-AST/29245/2017, UIDB/04434/2020 (DOI: 10.54499/UIDB/04434/2020) and UIDP/04434/2020 (DOI: 10.54499/UIDP/04434/2020) and UI/BD/152315/2021. 

The MeerKAT telescope is operated by the South African Radio Astronomy Observatory, which is a facility of the National Research Foundation, an agency of the Department of Science and Innovation. We acknowledge the use of the ilifu cloud computing facility – www.ilifu.ac.za, a partnership between the University of Cape Town, the University of the Western Cape, Stellenbosch University, Sol Plaatje University and the Cape Peninsula University of Technology. The Ilifu facility is supported by contributions from the Inter-University Institute for Data Intensive Astronomy (IDIA – a partnership between the University of Cape Town, the University of Pretoria and the University of the Western Cape, the Computational Biology division at UCT and the Data Intensive Research Initiative of South Africa (DIRISA). The authors acknowledge the Centre for High Performance Computing (CHPC), South Africa, for providing computational resources to this research project.
LOFAR is the Low Frequency Array, designed and constructed by ASTRON. It has observing, data processing, and data storage facilities in several countries, which are owned by various parties (each with their own funding sources), and which are collectively operated by the ILT foundation under a joint scientific policy. The ILT resources have benefited from the following recent major funding sources: CNRS-INSU, Observatoire de Paris and Université d'Orléans, France; BMBF, MIWF-NRW, MPG, Germany; Science Foundation Ireland (SFI), Department of Business, Enterprise and Innovation (DBEI), Ireland; NWO, The Netherlands; The Science and Technology Facilities Council, UK; Ministry of Science and Higher Education, Poland; The Istituto Nazionale di Astrofisica (INAF), Italy. 

This research made use of the University of Hertfordshire
high-performance computing facility (\url{https://uhhpc.herts.ac.uk}).

\section*{Data Availability}

The parent MIGHTEE catalogue is publicly available at \citealt{Haleetal2024}. LOFAR and GMRT data are available from public repositories\footnote{For LOFAR data see \url{https://lofar-surveys.org/}.}. Other data used to support the analysis in this paper are available on reasonable request to the corresponding author.



\bibliographystyle{mnras}
\bibliography{reference3} 




\appendix

\section{$\alpha$-z and $\alpha$-L for compact 3CRR sources}
\label{app:3csubset}
\begin{figure*}
	\includegraphics[width=7.0in,height=7.0in]{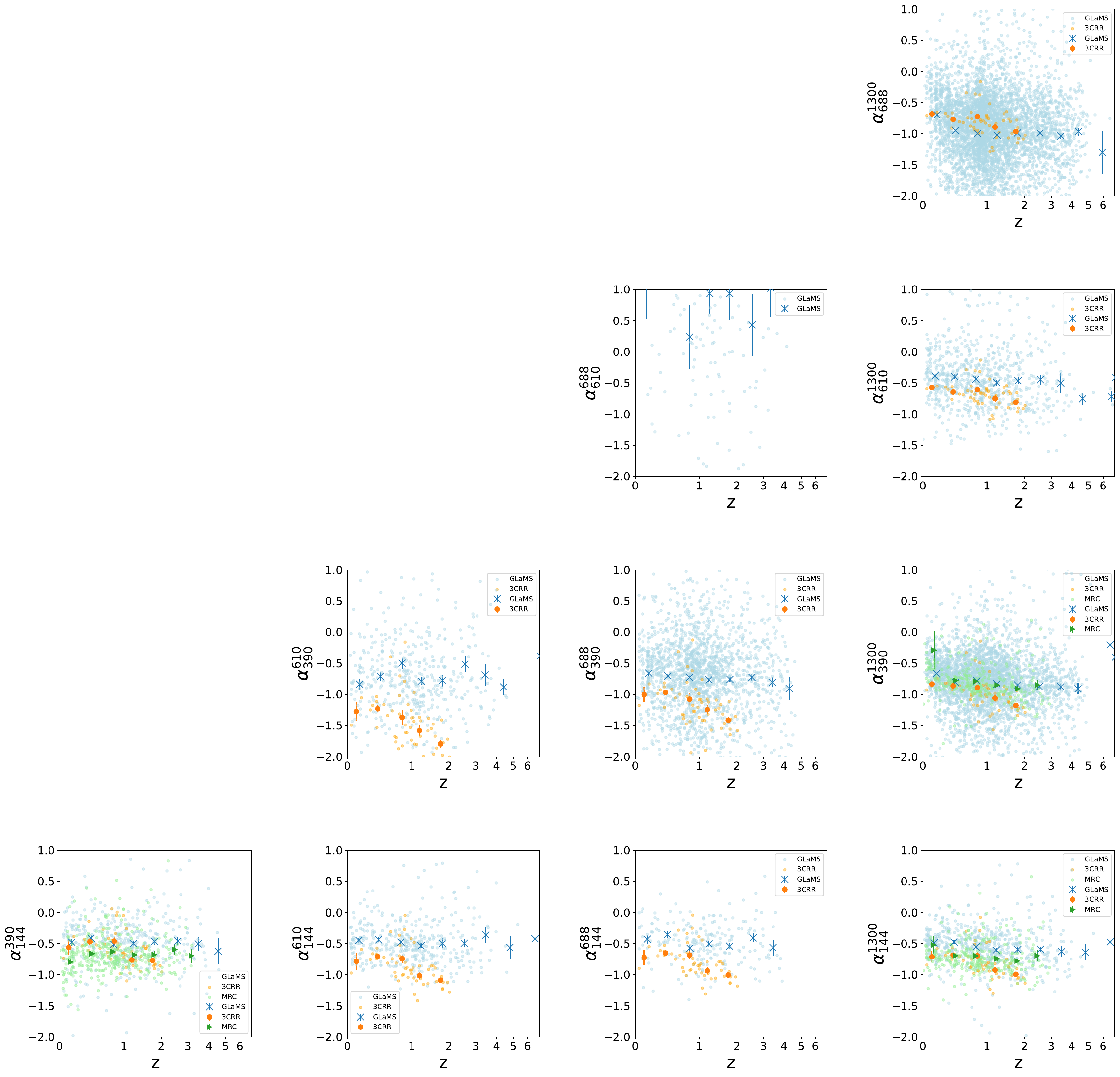}
	\caption{The correlation of the average spectral index with the redshift of the sources where the spectral index is calculated for flux densities measured at 144 MHz (LOFAR), 390 MHz (uGMRT band-3), 610 MHz (GMRT), 688 MHz (uGMRT band-4), and 1.3 GHz (MIGHTEE). Individual data points are also plotted without errors in order to indicate the spread of the data. The MRC sample is not present for some plots as the sample size for these frequencies is very low. Error bars indicate the $1\sigma$ error on the weighted mean. The sources in 3CRR sample are selected by filtering sources having angular size less than 10 arcsec}
	\label{2ALZ_3cless10}
\end{figure*}

\begin{figure*}
	\includegraphics[width=7.0in,height=7.0in]{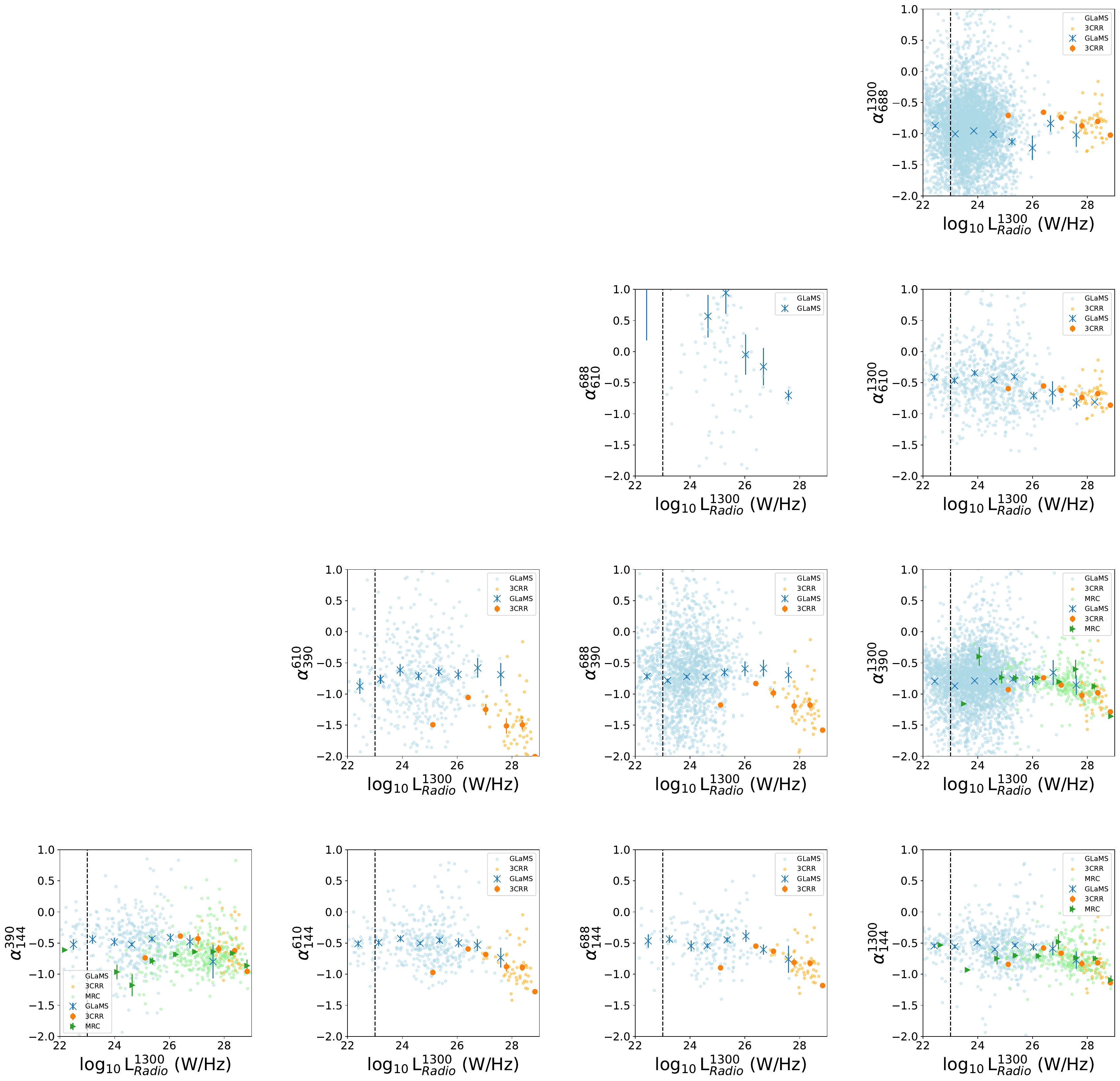}
	\caption{The correlation of the spectral index with the luminosity of the sources from MIGHTEE at 1.3 GHz where the spectral index is calculated for flux densities measured at 144 MHz (LOFAR), 390 MHz (uGMRT band-3), 610 MHz (GMRT), 688 MHz (uGMRT band-4), and 1.3 GHz (MIGHTEE). Luminosity is calculated as described in Section \ref{sec:PA}. Comments as in Fig.\ \ref{2ALZ}. The vertical dotted line shows the radio luminosity value above which AGN start to dominate. The sources in 3CRR sample are selected by filtering sources having angular size less than 10 arcsec}
	\label{2ALL_3cless10}
\end{figure*}

In figure \ref{2ALZ_3cless10} and \ref{2ALL_3cless10} we show the $\alpha$-$z$ and $\alpha$-L plots where the sample using the 3CRR data is also filtered for sources having largest angular sizes less than 20 arcsec (consistent with being unresolved with {\tt DC\_Maj} $<10$ arcsec since {\tt DC\_Maj} is the FWHM of a Gaussian). The number of sources after filtering for angular sizes is 55 which is around 61 per cent of the non filtered 3CRR sample used in the analysis above. We observe that this does not affect the overall position of the 3CRR sources with respect to the plots shown in the main body of the paper.

\bsp	
\label{lastpage}
\end{document}